\documentclass[nofootinbib]{revtex4-2}

\usepackage{geometry}
\usepackage{amsmath,amssymb,amsfonts,dcolumn,color,graphicx}
\usepackage{latexsym,placeins,epsfig,tikz,amsbsy,bm,enumitem}
\usetikzlibrary{arrows,shapes}
\usepackage{subfigure,rotating,bm,mathrsfs}
\usepackage[title]{appendix}
\usepackage[pagebackref=false,
colorlinks=true,linkcolor=black,citecolor=black,urlcolor=black]{hyperref}

\allowdisplaybreaks

\newcommand{\cH}{\mathcal H}

\newcommand{\ud}{\mathrm{UD}}
\newcommand{\lgauge}{\mathrm{L}}
\newcommand{\sfgauge}{\mathrm{SF}}
\newcommand{\cgauge}{\mathrm{C}}
\newcommand{\twoemt}{\mathrm{2EMT}}
\newcommand{\Order}{\mathcal O}
\newcommand{\factLW}{\mathcal F_{\mathrm{LW}}}
\newcommand{\factSW}{\mathcal F_{\mathrm{SW}}}

\begin{document}

\title{Quadratic effective energy-momentum tensor on uniform-density hypersurfaces during slow-roll inflation}

\author{Inyong Cho} \email{iycho@seoultech.ac.kr} \affiliation{School of Natural Sciences, College of Liberal Arts, Seoul National University of
Science and Technology, Seoul 01811, Republic of Korea}


\begin{abstract}

We investigate the quadratic-order effective energy-momentum tensor (2EMT) of scalar cosmological perturbations on uniform-density hypersurfaces
during slow-roll inflation.
The 2EMT is a gauge-fixed effective source rather than a gauge-invariant observable.
We impose the complete scalar gauge conditions $\delta\rho=0$ and $E=0$, express all perturbations in terms
of the Bardeen potential $\Psi$, and distinguish the strict infrared and ultraviolet limits from the
intermediate regimes. The uniform-density time shift differs from the comoving one by an explicitly gradient-suppressed
term, $Q_{\ud}-Q_{\cgauge}=-\Delta\Psi/[3{\mathcal H}({\mathcal H}'-{\mathcal H}^2)]$.
The two sources vanish at zero gradient, while their first nonzero infrared coefficients differ and
give $w_{\rm raw}\simeq-1/3$ at leading slow-roll order. In the intermediate infrared, all four gauges have
this leading ratio, with gauge-dependent amplitudes. In the ultraviolet, the longitudinal, spatially-flat, and comoving gauges
give $w_{\rm raw}\simeq1/3$, whereas the uniform-density gauge gives $w_{\rm raw}\simeq1$.
The common $1/\epsilon$ enhancement of the matter-clock gauges reflects the slow evolution of their clocks;
the additional uniform-density $1/\sigma_2^2$ enhancement originates from the Laplacian contribution to the density constraint. Neither
enhancement is interpreted as a gauge-independent physical instability, and the gauge reconstruction must remain perturbatively
small.

\end{abstract}

\maketitle

\section{Introduction}

The effective influence of cosmological perturbations on a homogeneous background first appears at quadratic order.
A frequently used description is the second-order effective energy-momentum tensor (2EMT), constructed from the terms
quadratic in the linear perturbations after the second-order Einstein equation is rearranged into a linear
equation for genuine second-order fields with a quadratic effective source.
This construction has been studied for scalar perturbations sourced by a perfect fluid and by
a canonical scalar field, as well as for coupled scalar and tensor perturbations during inflation
\cite{Cho:2020zbh,Cho:2022maa,Cho:2024vpv,Mukhanov:1996ak,Abramo:1997hu}, while related aspects of cosmological backreaction have also been investigated in Refs.~\cite{Geshnizjani:2002wp,Brandenberger:2002sk,Geshnizjani:2003cn,Martineau:2005aa}.

A central issue is gauge dependence.
The interpretation of the quadratic effective energy-momentum tensor and cosmological backreaction has long involved gauge-related
subtleties \cite{Unruh:1998ic,Ishibashi:2005sj,Green:2010qy}.
In our previous studies, explicit calculations showed that the gauge-fixed 2EMT depends on the choice
of slicing and threading \cite{Cho:2020zbh,Cho:2022maa,Cho:2024vpv}.
The earlier inflationary calculation considered the longitudinal, spatially-flat, and comoving gauges, while the later scalar--tensor
extension found especially strong gauge sensitivity in the scalar sector and motivated the examination of
additional gauge conditions.

The uniform-density gauge is a natural next case.
Its temporal hypersurfaces are selected by the matter clock $\rho(x)=\rho_0(\eta)$, rather than by vanishing shear,
spatial curvature, or inflaton fluctuation.
The use of matter fields as physical clocks in defining local cosmological quantities has been
emphasized in earlier studies of cosmological backreaction \cite{Geshnizjani:2002wp,Brandenberger:2002sk,Geshnizjani:2003cn}.
This gives the gauge a clear relational interpretation, but it does not make the raw
2EMT an observable.
A gauge-fixed quadratic source and a fully specified relational observable are different objects: the former
is definite after gauge fixing, whereas the latter additionally requires the same physical clock, observer,
hypersurface, and averaging prescription in every gauge.

The uniform-density gauge is also theoretically informative for two reasons.
First, for a single adiabatic scalar degree of freedom, {\it uniform-density} and {\it comoving} hypersurfaces
approach one another on super-Hubble scales.
This provides a nontrivial consistency check on the zero-gradient 2EMT, but does not require agreement
of the first finite-gradient coefficients. More broadly, the interpretation of a single-field inflationary background as
providing a preferred clock is standard in the effective-field-theory description of inflation \cite{Cheung:2007st}.
Second, during slow roll the background density evolves slowly, $\rho_0'\propto\epsilon$.
In the uniform-density gauge $\delta\rho_{\ud}=0$, while the gauge-invariant density perturbation satisfies $\delta\rho_{\rm gi}=-\rho_0'Q_{\ud}$, where $Q
\equiv\beta+E'$ is a gauge variable.
Hence $Q_{\ud}=-\delta\rho_{\rm gi}/\rho_0'$, so a slowly evolving density background makes the slicing increasingly sensitive to
a fixed gauge-invariant density perturbation.
This sensitivity is further amplified for short-wavelength modes because $\delta\rho_{\rm gi}$ contains the Laplacian contribution
proportional to $k^2\Psi$.
The resulting enhancement therefore exposes how the conditioning of a physical clock enters a gauge-fixed
quadratic source.

In this work we derive and analyze the 2EMT in the uniform-density gauge during slow-roll
inflation.
The longitudinal, spatially-flat, and comoving gauges were studied previously in Ref.~\cite{Cho:2022maa}.
Here we re-evaluate those three sectors directly from the general Fourier-space 2EMT, and present them
in a more explicit three-stage form: the full expression in $k$ and $\cH$, the equivalent
expression in $\sigma_1=k/\cH$ or $\sigma_2=\cH/k$ appropriate to the ``strict'' infrared or ultraviolet limit, $\sigma_i\to0$ at
fixed nonzero slow-roll parameters, and the slow-roll-reordered expression appropriate to the ``intermediate'' regime $\epsilon,|\delta|\ll\sigma_i^2\ll1$.
Thus the comparison formulas used in this work are not merely transcriptions of the abbreviated
expressions in Ref.~\cite{Cho:2022maa}.
This reorganization keeps the slow-roll and wavelength dependences separately visible, and refines the interpretation of
which terms dominate in the different asymptotic regimes.
We use \emph{exact} background relations expressing $\mathcal H'$, $\phi_0'$, and the potential and its derivatives,
$V$, $V_\phi$, and $V_{\phi\phi}$, in terms of $\mathcal H$ and the slow-roll parameters.
This differs from the lowest-order background treatment in Ref.~\cite{Cho:2022maa} and modifies some of its results
for the three gauges considered there.

The paper is organized as follows.
In Sec.~\ref{sec:setup} we review the scalar perturbations, gauge-invariant variables, linear solutions, and the definition of
the 2EMT.
In Sec.~\ref{sec:UD} we impose the uniform-density gauge and derive the corresponding clock displacement.
Sections~\ref{sec:UDLW} and \ref{sec:UDSW} present the uniform-density results in the strict and intermediate infrared and ultraviolet
regimes.
In Sec.~\ref{sec:comparison} we compare the four gauge conditions and identify the regime-dependent patterns.
Section~\ref{sec:interpretation} discusses the physical-clock interpretation and the limitations of the raw 2EMT.
We conclude in Sec.~\ref{sec:conclusion}.
The full recalculated results for the three comparison gauges are collected in the appendices.
The uniform-expansion gauge, which is closely related to constant-mean-curvature slicing in long-wavelength treatments \cite{Harada:2015yda}, will
be analyzed separately in a companion work.

\section{Scalar perturbations and the quadratic effective source}

\label{sec:setup}

\subsection{Metric, matter field, and gauge transformations}

We consider scalar perturbations of a spatially flat Friedmann background,
\begin{equation}
ds^2=a^2(\eta)\left\{-(1+2\alpha)d\eta^2-2\beta_{,i}d\eta\,dx^i
+\left[(1-2\psi)\delta_{ij}
+2E_{,ij}\right]dx^i dx^j\right\},
\label{eq:metric}
\end{equation}
and a canonical scalar field
\begin{equation}
\phi(\eta,\bm x)=\phi_0(\eta)+\delta\phi(\eta,\bm x).
\end{equation}
The energy-momentum tensor is
\begin{equation}
T_{\mu\nu}=\phi_{,\mu}\phi_{,\nu}-g_{\mu\nu}
\left[\frac12g^{\rho\sigma}\phi_{,\rho}\phi_{,\sigma}
+V(\phi)\right].
\end{equation}
The background equations are
\begin{align}
\cH^2=\frac{8\pi G}{3}\left(\frac12\phi_0'^2+a^2V\right),
\end{align}
\begin{align}
2\cH'+\cH^2=8\pi G\left(-\frac12\phi_0'^2+a^2V\right),
\end{align}
\begin{align}
\phi_0''+2\cH\phi_0'+a^2V_\phi=0,
\end{align}
where $\cH=a'/a$ and a prime denotes $d/d\eta$.

Under the infinitesimal transformation $x^\mu \to x^\mu + \xi^\mu$ generated by $\xi^\mu=(\xi^0,\partial^i\zeta)$, the perturbation variables
transform as
\begin{align}
\widetilde\alpha&=\alpha-\xi^{0\prime}-\cH\xi^0,\\
\widetilde\beta&=\beta-\xi^0+\zeta',\\
\widetilde\psi&=\psi+\cH\xi^0,\\
\widetilde E&=E-\zeta,\\
\widetilde{\delta\phi}&=\delta\phi-\phi_0'\xi^0.
\end{align}
Introducing
\begin{equation}
Q\equiv\beta+E',
\qquad
\widetilde Q=Q-\xi^0,
\end{equation}
we define the first-order gauge-invariant combinations in the standard scalar-perturbation framework \cite{Mukhanov:1990me}
\begin{align}
\Phi&=\alpha-Q'-\cH Q,\\
\Psi&=\psi+\cH Q,\\
\delta\phi_{\rm gi}&=\delta\phi-\phi_0'Q.
\end{align}
Here, $\Phi$ and $\Psi$ are the so called Bardeen potentials.
For a canonical scalar field without anisotropic stress, $\Phi=\Psi$.
The first-order equations give
\begin{align}
\delta\phi_{\rm gi}&=\frac{\Psi'+\cH\Psi}{4\pi G\phi_0'},\label{eq:dphigi}\\
\delta\phi_{\rm gi}'&=\frac{\Delta\Psi-K\Psi'-L\Psi}{4\pi G\phi_0'},
\end{align}
where
\begin{equation}
K=3\cH+a^2\frac{V_\phi}{\phi_0'},\qquad
L=\cH'+2\cH^2+a^2\cH\frac{V_\phi}{\phi_0'}.
\end{equation}
The Bardeen potential obeys
\begin{equation}
\Psi''-\Delta\Psi+2K\Psi'+2L\Psi=0.
\label{eq:PsiEOM}
\end{equation}
\subsection{Slow-roll and wavelength expansions}

We use the slow-roll parameters
\begin{equation}
\epsilon\equiv-\frac{\dot H}{H^2}=4\pi G\frac{\dot\phi_0^2}{H^2},
\qquad
\delta\equiv-\frac{\ddot\phi_0}{H\dot\phi_0}
=\epsilon-\frac{\dot\epsilon}{2H\epsilon},
\label{eq:slowroll}
\end{equation}
where $H=\dot a/a$ and an overdot denotes $d/dt$, with $t$ the cosmic time and $dt=a\,d\eta$.
In conformal time,
\begin{equation}
\cH'-\cH^2=-\epsilon\cH^2.
\label{eq:Hprime}
\end{equation}
We also introduce a dimensionless parameter that measures the third cosmic-time derivative of the background
inflaton:
\begingroup
\begin{equation}
\xi_3\equiv\frac{\dddot\phi_0}{H^2\dot\phi_0}.
\label{eq:xi3}
\end{equation}
\endgroup
It enters the exact evolution equation for $\delta$, $\dot\delta=H[\delta(\epsilon+\delta)-\xi_3]$, and the background relation for $V_{\phi\phi}$.
In the slow-roll hierarchy, $\xi_3$ is of second order, like $\epsilon^2$, $\epsilon\delta$, and $\delta^2$. The
background identities used in the symbolic reduction are
\begingroup
\begin{equation}
\begin{gathered}
\epsilon'=2\cH\epsilon(\epsilon-\delta),\qquad
\delta'=\cH[\delta(\epsilon+\delta)-\xi_3],\\
\cH'=(1-\epsilon)\cH^2,\qquad
\phi_0'=-\sqrt{2\epsilon}\,M_p\cH,\\
a^2V=(3-\epsilon)M_p^2\cH^2,\quad
a^2V_\phi=\sqrt{2\epsilon}(3-\delta)M_p\cH^2,\quad
a^2V_{\phi\phi}=\cH^2[3(\epsilon+\delta)-\xi_3],
\end{gathered}
\label{eq:bgconsistent}
\end{equation}
\endgroup
where $M_p^{-2}=8\pi G$.
Equivalently, $a^2V_{\phi\phi}/\cH^2=(3-\delta)(\epsilon+\delta)+\delta'/\cH$. Although $\xi_3$ is of second slow-roll order, we retain it in the exact
background relations.
Inverse powers of $\epsilon$ in the comoving and uniform-density gauges can promote its contributions to
lower orders in the final 2EMT.

We decompose the perturbation into Fourier modes using the convention
\begin{align}
\Psi(\eta,\bm x)
&=\int d^3k\,\Psi_{\bm k}(\eta)e^{i\bm k\cdot\bm x},
\label{eq:FourierExpansion}\\
\Psi_{\bm k}(\eta)
&=\frac{1}{(2\pi)^3}\int d^3x\,\Psi(\eta,\bm x)e^{-i\bm k\cdot\bm x}.
\label{eq:FourierTransform}
\end{align}
The first relation is the Fourier-mode expansion, whereas the second transforms the configuration-space field to
momentum space.
For each Fourier mode, the long- and short-wavelength solutions are written as \cite{Cho:2022maa,Cho:2024vpv}
\begin{equation}
\Psi_{\bm k}(\eta)\simeq
\begin{cases}
A_1\epsilon,
& \bigl(k\ll\cH:\ \text{long-wavelength limit}\bigr),\\[1mm]
\displaystyle 4\pi G\frac{\phi_0'}{a}
\left[c_1\sin(k\eta)+c_2\cos(k\eta)\right],
& \bigl(k\gg\cH:\ \text{short-wavelength limit}\bigr).
\end{cases}
\label{eq:PsiSolutions}
\end{equation}
For short-wavelength products we retain the slowly varying envelope when differentiating and only then take
the fast-phase average,
\begingroup
\begin{equation}
\overline{\sin^2(k\eta)}=\overline{\cos^2(k\eta)}=\frac12,
\qquad \overline{\sin(k\eta)\cos(k\eta)}=0.
\label{eq:phaseaverage}
\end{equation}
\endgroup
We introduce independent expansion parameters in the long- and short-wavelength domains respectively,
\begin{equation}
\sigma_1\equiv\frac{k}{\cH}\ll1,
\qquad
\sigma_2\equiv\frac{\cH}{k}\ll1.
\label{eq:sigmas}
\end{equation}
The slow-roll and gradient (wavelength) limits must be tracked separately.
Throughout this paper we distinguish four asymptotic regimes.
The \emph{strict IR} and \emph{strict UV} limits are obtained by expanding first in $\sigma_1$ or
$\sigma_2$, respectively, at fixed admissible nonzero slow-roll parameters, taking $\sigma_i\to0$. The \emph{intermediate IR} and \emph{intermediate
UV} regimes are defined by
\begin{equation}
\epsilon,|\delta|\ll\sigma_i^2\ll1,
\label{eq:intermediateHierarchy}
\end{equation}
and are organized by expanding first in the slow-roll parameters and then according to the
wavelength hierarchy.

We retain the explicit gradient terms ($k^2$, $k^4$, $\ldots$) in the 2EMT while using the
leading asymptotic mode functions for $\Psi$ in Eq.~\eqref{eq:PsiSolutions}.
Higher-order corrections to $\Psi$ are not included.

\subsection{Definition of the 2EMT}

At second perturbative order, the Einstein equation can be arranged as
\begin{equation}
G^{(1)}_{\mu\nu}[g^{(2)}]
=8\pi G T^{(1)}_{\mu\nu}[g^{(2)},\delta\phi^{(2)}]
+8\pi G T^{(2,\mathrm{eff})}_{\mu\nu},
\end{equation}
where
\begin{equation}
T^{(2,\mathrm{eff})}_{\mu\nu}
\equiv T^{(2)}_{\mu\nu}[g^{(1)},\delta\phi^{(1)}]
-\frac{1}{8\pi G}G^{(2)}_{\mu\nu}[g^{(1)}].
\label{eq:2EMTdef}
\end{equation}
The quadratic effective source is evaluated in Fourier space.
We denote the contribution associated with a fixed Fourier pair $\bm k$ and $-\bm k$
by
\begin{equation}
\widehat\tau_{\mu\nu}(k)\equiv8\pi G\tau_{\mu\nu}(k)
=8\pi G\left\langle T^{(2,\mathrm{eff})}_{\mu\nu}\right\rangle_k.
\label{eq:tauaverage}
\end{equation}
Here $\langle\cdots\rangle_k$ denotes the Fourier-transformed quadratic quantity obtained through the required spatial integration; it does
not represent a separate averaging prescription.
The integration pairs the modes $\bm k$ and $-\bm k$.
For a real perturbation, $\Psi_{-\bm k}=\Psi_{\bm k}^{*}$, and similarly for the other fields.
Below we suppress the explicit argument $k$ of the fixed-mode contribution.

The fixed Fourier pair is not itself isotropic. For the isotropic scalar contribution, we additionally
use the angular or statistically isotropic projection, following Abramo et al.~\cite{Abramo:1997hu},
\begingroup
\begin{equation}
k_i k_j\longrightarrow\frac{k^2}{3}\delta_{ij},\qquad
\widehat\tau_{ii}\equiv\frac13\delta^{ij}\widehat\tau_{ij}.
\label{eq:isoproj}
\end{equation}
\endgroup
This projection makes the off-diagonal spatial components vanish, $\widehat\tau_{ij}=0$ for $i\neq j$. The mixed components
also vanish, $\widehat\tau_{0i}=0$, because the isotropic angular average of $k_i$ is zero. The corresponding diagnostic
energy density, pressure, and ratio are
\begin{equation}
\rho_{\rm raw}=\frac{\tau_{00}}{a^2},\qquad
p_{\rm raw}=\frac{\tau_{ii}}{a^2},\qquad
w_{\rm raw}=\frac{p_{\rm raw}}{\rho_{\rm raw}}.
\label{eq:wraw}
\end{equation}
The label ``raw'' emphasizes that these gauge-fixed quantities are not, by themselves, relational observables.
In particular, $w_{\rm raw}$ is introduced only as a diagnostic component ratio and should not
be identified with a physical equation-of-state parameter without a common observer and relational prescription.

The general expression for $\tau_{\mu\nu}$, as presented in Appendix~\ref{app:general2EMT}, depends on the gauge-invariant potential $\Psi$
and on the gauge variables $Q$ and $E$ \cite{Cho:2022maa}.
A complete gauge condition fixes $Q$ and $E$, after which the result can be written
solely in terms of $\Psi$, but its functional form remains gauge dependent.

\section{Uniform-density gauge}

\label{sec:UD}

\subsection{Gauge conditions and clock displacement}

The temporal gauge freedom is fixed by the uniform-density condition $\delta\rho=0$, which selects hypersurfaces with
no first-order density perturbation.
The remaining scalar spatial gauge freedom is fixed independently by imposing $E=0$.
The complete gauge conditions are therefore
\begin{equation}
\delta\rho=0,\qquad E=0.
\label{eq:UDconditions}
\end{equation}
For the scalar field,
\begin{equation}
\delta\rho=\frac{1}{a^2}\left(\phi_0'\delta\phi'-\phi_0'^2\alpha\right)+V_\phi\delta\phi
=\frac{1}{a^2}\left[\phi_0'\delta\phi'-\phi_0'^2\alpha-(\phi_0''+2\cH\phi_0')\delta\phi\right].
\label{eq:drho}
\end{equation}
The gauge-invariant density perturbation is
\begin{equation}
\delta\rho_{\rm gi}=\delta\rho-\rho_0'Q.
\end{equation}
In the uniform-density gauge, $\delta\rho_{\ud}=0$, so this relation becomes
\begin{equation}
Q_{\ud}=-\frac{\delta\rho_{\rm gi}}{\rho_0'}.
\label{eq:QUDclock}
\end{equation}
This form makes the clock interpretation explicit: when $|\rho_0'|$ is small, a fixed gauge-invariant density
perturbation corresponds to a larger slicing displacement $Q_{\ud}$.
The $00$ component of the first-order equations can be written as
\begin{equation}
3\cH(\Psi'+\cH\Psi)-\Delta\Psi=-4\pi G a^2\delta\rho_{\rm gi}.
\label{eq:00constraint}
\end{equation}
Using $\delta\rho_{\ud}=0$, $\rho_0'=-3\cH\phi_0'^2/a^2$, and $4\pi G\phi_0'^2=\cH^2-\cH'$, we obtain
\begin{equation}
Q_{\ud}=\frac{3\cH(\Psi'+\cH\Psi)-\Delta\Psi}
{3\cH(\cH'-\cH^2)}.
\label{eq:QUD}
\end{equation}
Since $E=0$, $Q_{\ud}=\beta_{\ud}$.
The remaining perturbations are
\begin{align}
\beta_{\ud}&=Q_{\ud},\\
\psi_{\ud}&=\Psi-\cH Q_{\ud},\\
\alpha_{\ud}&=\Psi+Q_{\ud}'+\cH Q_{\ud},\\
\delta\phi_{\ud}&=\frac{\Psi'+\cH\Psi}{4\pi G\phi_0'}
+\phi_0'Q_{\ud}.
\label{eq:UDvariables}
\end{align}
For comparison, the comoving gauge has $\delta\phi=E=0$ and
\begin{equation}
Q_{\cgauge}=\frac{\Psi'+\cH\Psi}{\cH'-\cH^2}.
\label{eq:QC}
\end{equation}
Therefore, the difference becomes
\begin{equation}
Q_{\ud}-Q_{\cgauge}=-\frac{\Delta\Psi}{3\cH(\cH'-\cH^2)}.
\label{eq:Qdifference}
\end{equation}
Equation~\eqref{eq:Qdifference} is the key structural relation.
In the long-wavelength limit the two clocks agree up to $\Order(k^2/\cH^2)$.
In the short-wavelength limit the difference is enhanced by
\begin{equation}
Q_{\ud}-Q_{\cgauge}\sim \frac{1}{\epsilon\cH}\frac{k^2}{\cH^2}\Psi,
\label{eq:Qenhancement}
\end{equation}
up to an overall sign fixed by the Fourier convention.
The factor $1/\epsilon$ follows from the slow evolution of the density clock, $\rho_0'\propto\epsilon$, while the
additional factor $k^2/\cH^2$ comes from the Laplacian (gradient) term in $\delta\rho_{\rm gi}$.

\section{Uniform-density 2EMT in the long-wavelength limit}

\label{sec:UDLW}

Substitution of Eqs.~\eqref{eq:QUD}--\eqref{eq:UDvariables} into the general 2EMT gives the following long-wavelength expressions.
Here $\sigma_1 = k/\cH\ll1$.
The strict IR limit means $\sigma_1\to0$ at fixed nonzero slow-roll parameters, whereas the intermediate IR
regime is defined by $\epsilon,|\delta|\ll\sigma_1^2\ll1$.\footnote{We checked whether the omitted $\Order(k^2)$ correction
to the leading IR mode $\Psi_k$ in Eq.~\eqref{eq:PsiSolutions}
changes the first nonzero IR terms.
Using the linear mode equation and the exact background identities,
each density and pressure expression in the comoving and uniform-density gauges
is proportional to $k^2$,
while keeping $\Psi_k$ and $\Psi_k'$ unspecified.
For example, substituting
$\Psi_k=\Psi_0+k^2\Psi_2+\Order(k^4)$ and
$\Psi_k'=\Psi_0'+k^2\Psi_2'+\Order(k^4)$
shows that the corrections $\Psi_2$ and $\Psi_2'$
contribute only at $\Order(k^4)$,
because each 2EMT expression already carries a $k^2$ factor.
The first $k^2$ coefficients in
Eqs.~\eqref{eq:UDLW00new}, \eqref{eq:UDLWiin},
\eqref{eq:Clw00}, and \eqref{eq:Clwii}
are therefore unchanged.
This check applies to general backgrounds satisfying Eq.~\eqref{eq:bgconsistent}.}

We present the results of $\widehat{\tau}_{\mu\nu}$ in this paper in the following form.
The first line is the full expression obtained after inserting the gauge conditions, the mode
solution, and the slow-roll dependence of the background and perturbation quantities (including $\Phi=\Psi$, $\cH$, $\phi_0$,
$V$, and their derivatives) into the general 2EMT.
The second line is an exact algebraic rearrangement of that same expression in powers of
$\sigma_i$; it is not a truncation of a wavelength expansion, and it is the form
used to identify the strict IR/UV limit.
The third line reorganizes the result in the slow-roll parameters according to $\epsilon,|\delta|\ll\sigma_i^2\ll1$, and retains
the zeroth- and first-order terms in slow roll in each $\sigma_i$ coefficient after extracting the
displayed overall $\epsilon$ or $1/\epsilon$ factor. \footnote{For the longitudinal, spatially-flat and comoving gauges in Appendices,
and the uniform-density gauge in this and next sections, the full and $\sigma_i$-ordered forms are
polynomial or Laurent-polynomial in $\sigma_i$, so this rearrangement is algebraically exact.
The hierarchy of dominant terms nevertheless depends on the relative ordering of $\epsilon$, $\delta$, and
$\sigma_i^2$, which is why the truncated third representation is useful.} The results are as follows.

We count $\xi_3$ as second order in slow roll. The remainder $\Order_{\rm SR}(2;\sigma_i)$ denotes omitted
second-and-higher slow-roll coefficients at the wavelength powers generated by the adopted mode input; the overall
$1/\epsilon$, where present, remains outside this remainder. We use $\factLW\equiv\cH^2|A_1|^2$.
\begingroup
\begin{align}
\widehat\tau_{00}^{\mathrm{LW}}={}&\frac{|A_1|^2}{9 \cH^{4}}\Big\{k^{6}[12+(-6 \epsilon)]
+k^{4}\cH^{2}[-93+(280 \epsilon-84 \delta)+(-141 \epsilon^{2}
+35 \delta \epsilon)\nonumber\\
&\hspace{18mm}
+(9 \epsilon^{3}
+6 \delta \epsilon^{2}
+\epsilon \delta^{2}-\epsilon \xi_3)]
+k^{2}\cH^{4}[-162+(-369 \epsilon+288 \delta)+(369 \epsilon^{2}-324 \delta \epsilon+72 \delta^{2})\nonumber\\
&\hspace{18mm}
+(-36 \epsilon^{3}
+72 \delta \epsilon^{2}-36 \epsilon \delta^{2})]\Big\}
\nonumber\\
={}&|A_1|^2 \cH^{2}\Big\{[-18+(-41 \epsilon+32 \delta)+(41 \epsilon^{2}-36 \delta \epsilon+8 \delta^{2})+(-4 \epsilon^{3}
+8 \delta \epsilon^{2}-4 \epsilon \delta^{2})]\sigma_1^{2}\nonumber\\
&\hspace{18mm}
+[-\frac{31}{3}
+(\frac{280 \epsilon}{9}-\frac{28 \delta}{3})+(-\frac{47 \epsilon^{2}}{3}
+\frac{35 \delta \epsilon}{9})+(\epsilon^{3}
+\frac{2 \delta \epsilon^{2}}{3}
+\frac{\epsilon \delta^{2}}{9}-\frac{\epsilon \xi_3}{9})]\sigma_1^{4}\nonumber\\
&\hspace{18mm}
+[\frac{4}{3}
+(-\frac{2 \epsilon}{3})]\sigma_1^{6}\Big\}\;\text{(strict IR)}
\nonumber\\
\simeq{}&|A_1|^2 \cH^{2}\Big\{[-18\sigma_1^{2}-\frac{31}{3}\sigma_1^{4}
+\frac{4}{3}\sigma_1^{6}]\nonumber\\
&\hspace{18mm}
+[(-41 \epsilon+32 \delta)\sigma_1^{2}
+(\frac{280 \epsilon}{9}-\frac{28 \delta}{3})\sigma_1^{4}-\frac{2 \epsilon}{3}\sigma_1^{6}]
+\Order_{\rm SR}(2;\sigma_1)\Big\}\;\text{(intermediate IR)}.
\label{eq:UDLW00new}
\end{align}
\endgroup
\begingroup
\begin{align}
\widehat\tau_{ii}^{\mathrm{LW}}={}&\frac{|A_1|^2}{27 \cH^{6}}\Big\{-2k^{8}
+k^{6}\cH^{2}[114+(-132 \epsilon+12 \delta)+(30 \epsilon^{2}
+4 \delta \epsilon+4 \delta^{2}-4 \xi_3)]
+k^{4}\cH^{4}[-201\nonumber\\
&\hspace{18mm}
+(744 \epsilon-228 \delta)+(-633 \epsilon^{2}
+159 \delta \epsilon-72 \delta^{2}
+72 \xi_3)\nonumber\\
&\hspace{18mm}
+(243 \epsilon^{3}-54 \delta \epsilon^{2}
+51 \epsilon \delta^{2}-69 \epsilon \xi_3-24 \delta^{3}
+24 \delta \xi_3)]
+k^{2}\cH^{6}[162+(567 \epsilon-288 \delta)\nonumber\\
&\hspace{18mm}
+(297 \epsilon^{2}-252 \delta \epsilon-144 \delta^{2}
+72 \xi_3)+(108 \epsilon^{3}-648 \delta \epsilon^{2}
+180 \epsilon \delta^{2}
+216 \epsilon \xi_3+144 \delta^{3}-144 \delta \xi_3)]\Big\}
\nonumber\displaybreak[4]\\
={}&|A_1|^2 \cH^{2}\Big\{[6+(21 \epsilon-\frac{32 \delta}{3})+(11 \epsilon^{2}-\frac{28 \delta \epsilon}{3}-\frac{16 \delta^{2}}{3}
+\frac{8 \xi_3}{3})\nonumber\\
&\hspace{18mm}
+(4 \epsilon^{3}-24 \delta \epsilon^{2}
+\frac{20 \epsilon \delta^{2}}{3}
+8 \epsilon \xi_3+\frac{16 \delta^{3}}{3}-\frac{16 \delta \xi_3}{3})]\sigma_1^{2}
+[-\frac{67}{9}
+(\frac{248 \epsilon}{9}-\frac{76 \delta}{9})\nonumber\\
&\hspace{18mm}
+(-\frac{211 \epsilon^{2}}{9}
+\frac{53 \delta \epsilon}{9}-\frac{8 \delta^{2}}{3}
+\frac{8 \xi_3}{3})+(9 \epsilon^{3}-2 \delta \epsilon^{2}
+\frac{17 \epsilon \delta^{2}}{9}-\frac{23 \epsilon \xi_3}{9}-\frac{8 \delta^{3}}{9}
+\frac{8 \delta \xi_3}{9})]\sigma_1^{4}\nonumber\\
&\hspace{18mm}
+[\frac{38}{9}
+(-\frac{44 \epsilon}{9}
+\frac{4 \delta}{9})+(\frac{10 \epsilon^{2}}{9}
+\frac{4 \delta \epsilon}{27}
+\frac{4 \delta^{2}}{27}-\frac{4 \xi_3}{27})]\sigma_1^{6}-\frac{2}{27}\sigma_1^{8}\Big\}\;\text{(strict IR)}
\nonumber\\
\simeq{}&|A_1|^2 \cH^{2}\Big\{[6\sigma_1^{2}-\frac{67}{9}\sigma_1^{4}
+\frac{38}{9}\sigma_1^{6}-\frac{2}{27}\sigma_1^{8}]\nonumber\\
&\hspace{18mm}
+[(21 \epsilon-\frac{32 \delta}{3})\sigma_1^{2}
+(\frac{248 \epsilon}{9}-\frac{76 \delta}{9})\sigma_1^{4}
+(-\frac{44 \epsilon}{9}
+\frac{4 \delta}{9})\sigma_1^{6}]
+\Order_{\rm SR}(2;\sigma_1)\Big\}\;\text{(intermediate IR)}.
\label{eq:UDLWiin}
\end{align}
\endgroup
\subsection{Strict IR}

There is no $\sigma_1^0$ term in either component. Hence
\begingroup
\begin{equation}
\widehat\tau_{00}^{\ud},\widehat\tau_{ii}^{\ud}\longrightarrow0\qquad(\sigma_1\to0),
\label{eq:UDstrictIRzero}
\end{equation}
\endgroup
This satisfies the homogeneous constant-$\epsilon$ benchmark, but the strict-IR equation of state must be read
from the first nonzero gradient order rather than from the values at $k=0$. At fixed
admissible nonzero slow-roll parameters, Eqs.~\eqref{eq:UDLW00new} and \eqref{eq:UDLWiin} give
\begingroup
\begin{align}
\frac{\widehat\tau_{00}^{\ud}}{\factLW}&=-\sigma_1^2\Big\{18+(41\epsilon-32\delta)+\Order_{\rm SR}(2)\Big\}
+\Order(\sigma_1^4),\nonumber\\*
\frac{\widehat\tau_{ii}^{\ud}}{\factLW}&=\frac{\sigma_1^2}{3}\Big\{18+(63\epsilon-32\delta)+\Order_{\rm SR}(2)\Big\}
+\Order(\sigma_1^4),\nonumber\\*
\lim_{\sigma_1\to0}w_{\rm raw}^{\ud}&=-\frac13-\frac{11\epsilon}{27}
+\Order_{\rm SR}(2).
\label{eq:UDstrictIRratio}
\end{align}
\endgroup
Thus the leading slow-roll ratio in the strict IR is $-1/3$, even though both components
vanish. Only the ratio evaluated at exactly $k=0$ is the undefined expression $0/0$; this is
distinct from its finite limit.

\subsection{Intermediate IR}

Under $\epsilon,|\delta|\ll\sigma_1^2\ll1$, their leading nonzero terms are
\begingroup
\begin{equation}
\frac{\widehat\tau_{00}^{\ud}}{\factLW}\simeq-18\sigma_1^2,
\qquad
\frac{\widehat\tau_{ii}^{\ud}}{\factLW}\simeq6\sigma_1^2,
\qquad w_{\rm raw}^{\ud}\simeq-\frac13.
\label{eq:UDIntermediateIRleading}
\end{equation}
\endgroup
The third representations in Eqs.~\eqref{eq:UDLW00new} and \eqref{eq:UDLWiin} additionally display the $\sigma_1^4$, $\sigma_1^6$, and, for the
pressure, $\sigma_1^8$ terms with the retained linear slow-roll corrections. Compared with the comoving coefficients $(-10,10/3)$,
the uniform-density coefficients $(-18,6)$ expose a finite-gradient difference even though both sources vanish at $k=0$.

\section{Uniform-density 2EMT in the short-wavelength limit}

\label{sec:UDSW}

For the short-wavelength domain we use $\sigma_2\equiv\cH/k\ll1$.
The strict UV limit means $\sigma_2\to0$ at fixed nonzero slow-roll parameters, whereas the intermediate UV
regime is defined by $\epsilon,|\delta|\ll\sigma_2^2\ll1$.
As for the long-wavelength limit, the first line of each component is the full expression
after substituting the slow-roll dependence of the background and perturbation quantities.\footnote{Since $\Psi$ is sinusoidal in
the short-wavelength regime, as in Eq.~\eqref{eq:PsiSolutions}, the quadratic 2EMT components use the adiabatic fast-phase average
in Eq.~\eqref{eq:phaseaverage}, retaining the slowly varying envelope during differentiation.} The second line is its exact
algebraic rearrangement in powers of $\sigma_2$, with no wavelength truncation, and the third line keeps
the leading terms of the slow-roll-reordered form appropriate to the intermediate UV hierarchy.
The results are as follows. We write $C_k\equiv|c_1|^2+|c_2|^2$.
\begingroup
\begin{align}
\widehat\tau_{00}^{\mathrm{SW}}={}&\frac{2 \pi C_k G}{9 \cH^{2} a^{2} \epsilon}\Big\{k^{6}[3+(-\epsilon-2 \delta)]
+k^{4}\cH^{2}[-39+(115 \epsilon-6 \delta)+(-25 \epsilon^{2}-29 \delta \epsilon+3 \delta^{2})\nonumber\\
&\hspace{18mm}
+(\epsilon^{3}
+4 \delta \epsilon^{2}
+4 \epsilon \delta^{2}-\epsilon \xi_3)]
+k^{2}\cH^{4}[-162+(-153 \epsilon+180 \delta)+(45 \epsilon^{2}
+18 \delta \epsilon-18 \delta^{2})+(-9 \epsilon \delta^{2})]\Big\}
\nonumber\\
={}&\frac{2 \pi C_k G \cH^{4}}{a^{2} \epsilon}\Big\{[\frac{1}{3}
+(-\frac{\epsilon}{9}-\frac{2 \delta}{9})]\sigma_2^{-6}
+[-\frac{13}{3}
+(\frac{115 \epsilon}{9}-\frac{2 \delta}{3})+(-\frac{25 \epsilon^{2}}{9}-\frac{29 \delta \epsilon}{9}
+\frac{\delta^{2}}{3})\nonumber\\
&\hspace{18mm}
+(\frac{\epsilon^{3}}{9}
+\frac{4 \delta \epsilon^{2}}{9}
+\frac{4 \epsilon \delta^{2}}{9}-\frac{\epsilon \xi_3}{9})]\sigma_2^{-4}
+[-18+(-17 \epsilon+20 \delta)+(5 \epsilon^{2}
+2 \delta \epsilon-2 \delta^{2})\nonumber\\
&\hspace{18mm}
+(-\epsilon \delta^{2})]\sigma_2^{-2}\Big\}\;\text{(strict UV)}
\nonumber\\
\simeq{}&\frac{2 \pi C_k G \cH^{4}}{a^{2} \epsilon}\Big\{[\frac{1}{3}\sigma_2^{-6}-\frac{13}{3}\sigma_2^{-4}-18\sigma_2^{-2}]\nonumber\\
&\hspace{18mm}
+[(-\frac{\epsilon}{9}-\frac{2 \delta}{9})\sigma_2^{-6}
+(\frac{115 \epsilon}{9}-\frac{2 \delta}{3})\sigma_2^{-4}
+(-17 \epsilon+20 \delta)\sigma_2^{-2}]
+\Order_{\rm SR}(2;\sigma_2)\Big\}\;\text{(intermediate UV)}.
\label{eq:UDSW00new}
\end{align}
\endgroup
\begingroup
\begin{align}
\widehat\tau_{ii}^{\mathrm{SW}}={}&\frac{2 \pi C_k G}{27 \cH^{2} a^{2} \epsilon}\Big\{k^{6}[9+(-5 \epsilon-14 \delta)+(6 \epsilon^{2}
+4 \delta \epsilon+10 \delta^{2}-4 \xi_3)]
+k^{4}\cH^{2}[-39+(-129 \epsilon+150 \delta)\nonumber\\
&\hspace{18mm}
+(-21 \epsilon^{2}-9 \delta \epsilon-141 \delta^{2}
+72 \xi_3)+(27 \epsilon^{3}-12 \delta \epsilon^{2}
+36 \epsilon \delta^{2}-21 \epsilon \xi_3)]
+k^{2}\cH^{4}[162+(-81 \epsilon+36 \delta)\nonumber\\
&\hspace{18mm}
+(-243 \epsilon^{2}
+414 \delta \epsilon-342 \delta^{2}
+72 \xi_3)+(-117 \epsilon \delta^{2}
+72 \epsilon \xi_3+144 \delta^{3}-72 \delta \xi_3)]\Big\}
\nonumber\\
={}&\frac{2 \pi C_k G \cH^{4}}{3 a^{2} \epsilon}\Big\{[1+(-\frac{5 \epsilon}{9}-\frac{14 \delta}{9})+(\frac{2 \epsilon^{2}}{3}
+\frac{4 \delta \epsilon}{9}
+\frac{10 \delta^{2}}{9}-\frac{4 \xi_3}{9})]\sigma_2^{-6}
+[-\frac{13}{3}
+(-\frac{43 \epsilon}{3}
+\frac{50 \delta}{3})\nonumber\\
&\hspace{18mm}
+(-\frac{7 \epsilon^{2}}{3}-\delta \epsilon-\frac{47 \delta^{2}}{3}
+8 \xi_3)+(3 \epsilon^{3}-\frac{4 \delta \epsilon^{2}}{3}
+4 \epsilon \delta^{2}-\frac{7 \epsilon \xi_3}{3})]\sigma_2^{-4}
+[18+(-9 \epsilon+4 \delta)\nonumber\\
&\hspace{18mm}
+(-27 \epsilon^{2}
+46 \delta \epsilon-38 \delta^{2}
+8 \xi_3)+(-13 \epsilon \delta^{2}
+8 \epsilon \xi_3+16 \delta^{3}-8 \delta \xi_3)]\sigma_2^{-2}\Big\}\;\text{(strict UV)}
\nonumber\\
\simeq{}&\frac{2 \pi C_k G \cH^{4}}{3 a^{2} \epsilon}\Big\{[\sigma_2^{-6}-\frac{13}{3}\sigma_2^{-4}
+18\sigma_2^{-2}]\nonumber\\
&\hspace{18mm}
+[(-\frac{5 \epsilon}{9}-\frac{14 \delta}{9})\sigma_2^{-6}
+(-\frac{43 \epsilon}{3}
+\frac{50 \delta}{3})\sigma_2^{-4}
+(-9 \epsilon+4 \delta)\sigma_2^{-2}]
+\Order_{\rm SR}(2;\sigma_2)\Big\}\;\text{(intermediate UV)}.
\label{eq:UDSWiin}
\end{align}
\endgroup

For comparison among the gauges considered below, it is useful to factor out the common
$k^4$ short-wavelength derivative co-factor present in all four of the gauge-fixed results.
We therefore define
\begin{equation}
\factSW\equiv\frac{2\pi Gk^4}{a^2}
(|c_1|^2+|c_2|^2).
\label{eq:FSWdef}
\end{equation}
Since $k^4=\cH^4\sigma_2^{-4}$, $\factSW$ contains the common $\sigma_2^{-4}$ short-wavelength scaling.
Factoring out this common contribution does not remove ultraviolet momentum dependence from the 2EMT.
Rather, it provides a common baseline for comparing additional gauge-dependent enhancements.

\subsection{Strict UV}

The strict UV leading terms are therefore
\begingroup
\begin{align}
\frac{\widehat\tau_{00}^{\ud}}{\factSW}&\simeq\frac{3+(-\epsilon-2\delta)}{9\epsilon\sigma_2^2},\label{eq:UDstrictUV00}\\
\frac{\widehat\tau_{ii}^{\ud}}{\factSW}&\simeq
\frac{9+(-5\epsilon-14\delta)+(6\epsilon^2+4\epsilon\delta+10\delta^2-4\xi_3)}{27\epsilon\sigma_2^2}.
\label{eq:UDstrictUVii}
\end{align}
\endgroup
At leading slow-roll-independent numerator order,
\begingroup
\begin{equation}
\frac{\widehat\tau_{00}^{\ud}}{\factSW}\simeq\frac{1}{3\epsilon\sigma_2^2},\qquad
\frac{\widehat\tau_{ii}^{\ud}}{\factSW}\simeq\frac{1}{3\epsilon\sigma_2^2},\qquad
w_{\rm raw}^{\ud}\simeq1.
\label{eq:UDstrictUVw}
\end{equation}
\endgroup
The leading raw energy density is positive, and the leading raw pressure is also positive.
Restoring the common factor contained in $\factSW$ gives
\begin{equation}
\widehat\tau_{\mu\nu}^{\ud}\propto
\epsilon^{-1}\sigma_2^{-6},
\end{equation}
confirming that $\sigma_2^{-6}$ is the full strict-UV scaling of the uniform-density 2EMT.

The remaining factor after the common short-wavelength co-factor is removed can be written as
\begin{equation}
\frac{1}{\epsilon\sigma_2^2}
=\frac{1}{\epsilon}\frac{k^2}{\cH^2}.
\label{eq:UDUVenh}
\end{equation}
The full strict-UV uniform-density result therefore contains three conceptually distinct factors: the common $\sigma_2^{-4}$ short-wavelength
derivative scaling contained in $\factSW$, the $1/\epsilon$ matter-clock enhancement, and the additional $\sigma_2^{-2}$ gradient enhancement
specific to the uniform-density slicing.

The $1/\epsilon$ factor is also present in the comoving gauge, and reflects the poor conditioning
of the slowly evolving matter clock as the inflationary background approaches the de Sitter limit:
$\phi_0'\to0$ for the comoving clock and $\rho_0'\to0$ for the density clock.
The additional factor $\sigma_2^{-2}=k^2/\cH^2$ is specific to the uniform-density slicing, and originates from the Laplacian
term in the density constraint, Eq.~\eqref{eq:Qenhancement}.
These enhancements should not be interpreted as gauge-independent ultraviolet instabilities.

\subsection{Intermediate UV}

The third representations above give the explicit intermediate Laurent polynomials. In units of $\factSW$ the
leading slow-roll-independent numerator terms, including the first subleading wavelength contributions, are
\begingroup
\begin{align}
\frac{\widehat\tau_{00}^{\ud}}{\factSW}
&\simeq\frac{1}{3\epsilon}\left[\sigma_2^{-2}-13-54\sigma_2^2\right],
\label{eq:UDIntermediateUV00}\\
\frac{\widehat\tau_{ii}^{\ud}}{\factSW}
&\simeq\frac{1}{3\epsilon}\left[\sigma_2^{-2}-\frac{13}{3}
+18\sigma_2^2\right].
\label{eq:UDIntermediateUVii}
\end{align}
\endgroup
The linear slow-roll corrections to every displayed wavelength coefficient are retained in Eqs.~\eqref{eq:UDSW00new} and \eqref{eq:UDSWiin};
they must be included when accuracy beyond the zeroth-order numerator terms is required. The leading
intermediate ratio remains $w_{\rm raw}^{\ud}\simeq1$.

\subsection{Perturbative validity} The ultraviolet shift behaves as
\begingroup
\begin{equation}
Q_{\ud}\simeq-\frac{k^2}{3\epsilon\cH^3}\Psi,
\end{equation}
\endgroup
so reconstructed perturbations scale parametrically as
\begingroup
\begin{equation}
|\psi_{\ud}|\sim\frac{|\Psi|}{3\epsilon}\left(\frac{k}{\cH}\right)^2,
\qquad
|\alpha_{\ud}|,|k\beta_{\ud}|\sim\frac{|\Psi|}{3\epsilon}\left(\frac{k}{\cH}\right)^3.
\end{equation}
\endgroup
A necessary \emph{small-metric condition} is therefore
\begingroup
\begin{equation}
\frac{|\Psi|}{\epsilon}\left(\frac{k}{\cH}\right)^3\ll1.
\label{eq:UDvalidity}
\end{equation}
\endgroup
This condition applies to both UV regimes.
At fixed nonzero $|\Psi|$ and fixed $\epsilon$, it requires
\begingroup
\begin{equation}
\left(\frac{|\Psi|}{\epsilon}\right)^{1/3}\ll\sigma_2\ll1.
\label{eq:UDvalidityWindow}
\end{equation}
\endgroup
Thus, $\sigma_2$ cannot be made arbitrarily small while maintaining perturbative validity.
The strict UV limit $\sigma_2\to0$ defines a formal asymptotic coefficient.
Its application at finite wavelength requires the condition above. Applying the same analysis to the
reconstructed perturbations in the other three gauges gives the conditions summarized in Table~\ref{tab:UVvalidity}.

\begin{table}[!ht]
\begingroup
\small \setlength{\tabcolsep}{12pt}
\begin{ruledtabular}
\begin{tabular}{lll}
Gauge & UV small-metric condition & Allowed range of $\sigma_2$\\
\hline Uniform density & $|\Psi|/(\epsilon\sigma_2^3)\ll1$ &
$(|\Psi|/\epsilon)^{1/3}\ll\sigma_2\ll1$\\[2pt]  Longitudinal & $|\Psi|\ll1$ & $0<\sigma_2\ll1$ (no additional metric lower bound)\\[2pt] Spatially flat &
$|\Psi|/\sigma_2\ll1$ & $|\Psi|\ll\sigma_2\ll1$\\[2pt] Comoving & $|\Psi|/(\epsilon\sigma_2^2)\ll1$ & $(|\Psi|/\epsilon)^{1/2}\ll\sigma_2\ll1$
\end{tabular}
\end{ruledtabular}
\endgroup
\caption{Necessary UV small-metric conditions and their wavelength ranges at fixed oscillation amplitude $|\Psi|$ and fixed
background parameters.
These conditions do not by themselves establish the validity of the full perturbation expansion.}
\label{tab:UVvalidity}
\end{table}
\FloatBarrier

In the intermediate UV regime, this condition~\eqref{eq:UDvalidityWindow} must hold together with $\epsilon,|\delta|\ll\sigma_2^2\ll1$.
The allowed range therefore satisfies
\begingroup
\begin{equation}
\max\left\{\sqrt{\epsilon},\sqrt{|\delta|},
\left(\frac{|\Psi|}{\epsilon}\right)^{1/3}\right\}\ll\sigma_2\ll1.
\label{eq:UDintermediateValidity}
\end{equation}
\endgroup
For the other gauges, the intermediate UV regime likewise requires $\sqrt{\epsilon},\sqrt{|\delta|}\ll\sigma_2$ in addition to the
corresponding conditions in Table~\ref{tab:UVvalidity}.

The small-metric condition is necessary.
It does not by itself establish the validity of the full perturbation expansion or imply
a gauge-independent instability.  The $1/\epsilon$ enhancement of the 2EMT is identified by comparing its
$\epsilon$ dependence at fixed $c_i$, $a$, and $\mathcal H$.
Equation~\eqref{eq:PsiSolutions} gives $|\Psi_k|^2\propto\epsilon C_k$ at fixed background scales.
If $|\Psi_k|$ is held fixed instead, then $C_k\propto1/\epsilon$.
Both choices are valid, but they describe different normalizations of the perturbations.
Here we hold $c_i$, $a$, and $\mathcal H$ fixed.

\section{Comparison with longitudinal, spatially-flat, and comoving gauges}

\label{sec:comparison}

The longitudinal, spatially-flat, and comoving gauges were analyzed in Ref.~\cite{Cho:2022maa}.
For the present comparison, however, all three have been recalculated from the general Fourier-space 2EMT,
and are displayed explicitly in Appendices~\ref{app:L}--\ref{app:C} in the same three-stage form used above.
This representation is more detailed than the earlier presentation, and separates the strict and intermediate
orderings before physical interpretation.
The results are summarized in Table~\ref{tab:LWsummary} and~\ref{tab:SWsummary}.

\begin{table}[!ht]
\scriptsize \setlength{\tabcolsep}{3pt}
\begin{ruledtabular}
\begin{tabular}{lccc|ccc}
& \multicolumn{3}{c|}{Strict IR} & \multicolumn{3}{c}{Intermediate IR}\\ Gauge & $\widehat\tau_{00}/\factLW$ & $\widehat\tau_{ii}/\factLW$ & $w_{\rm raw}$ &
$\widehat\tau_{00}/\factLW$ & $\widehat\tau_{ii}/\factLW$ & $w_{\rm raw}$\\ \hline Longitudinal & $-(9\epsilon^2-3\epsilon\delta)$ & $(9\epsilon^2-3\epsilon\delta)$ & $-1$ &
$\epsilon\sigma_1^2$ & $-\frac13\epsilon\sigma_1^2$ & $-1/3$\\ Spatially flat & $-(9\epsilon^2-3\epsilon\delta)$ & $(9\epsilon^2-3\epsilon\delta)$ & $-1$ & $\epsilon\sigma_1^2$
& $-\frac13\epsilon\sigma_1^2$ & $-1/3$\\ Comoving & $-10\sigma_1^2$ & $\frac{10}{3}\sigma_1^2$ & $-1/3$ & $-10\sigma_1^2-\sigma_1^4+\cdots$ & $\frac{10}{3}\sigma_1^2-3\sigma_1^4+\cdots$
& $-1/3$\\ Uniform density & $-18\sigma_1^2$ & $6\sigma_1^2$ & $-1/3$ & $-18\sigma_1^2-\frac{31}{3}\sigma_1^4+\cdots$ & $6\sigma_1^2-\frac{67}{9}\sigma_1^4+\cdots$ &
$-1/3$
\end{tabular}
\end{ruledtabular}
\caption{Gauge-fixed scalar 2EMT in the long-wavelength domain, comparing the strict IR and intermediate IR regimes.
The strict-IR entries retain the first nonzero gradient order for the comoving and uniform-density gauges,
and $w_{\rm raw}$ denotes the ratio limit as $\sigma_1\to0$. For the comoving and uniform-density gauges,
the first finite-gradient corrections are shown in the intermediate-IR columns to make their separation away
from the strict limit explicit.
The displayed $w_{\rm raw}$ values are the leading diagnostic ratios.}
\label{tab:LWsummary}

\end{table}

\begin{table}[!ht]
\scriptsize \setlength{\tabcolsep}{3pt}
\begin{ruledtabular}
\begin{tabular}{lccc|ccc}
& \multicolumn{3}{c|}{Strict UV} & \multicolumn{3}{c}{Intermediate UV}\\ Gauge & $\widehat\tau_{00}/\factSW$ & $\widehat\tau_{ii}/\factSW$ & $w_{\rm raw}$ &
$\widehat\tau_{00}/\factSW$ & $\widehat\tau_{ii}/\factSW$ & $w_{\rm raw}$\\ \hline Longitudinal & $2$ & $2/3$ & $1/3$ &
$2+\sigma_2^2+\cdots$ & $\frac23-\frac13\sigma_2^2+\cdots$ & $1/3$\\ Spatially flat & $2$ & $2/3$ & $1/3$ & $2+\sigma_2^2+\cdots$
& $\frac23-\frac13\sigma_2^2+\cdots$ & $1/3$\\ Comoving & $1/\epsilon$ & $1/(3\epsilon)$ & $1/3$ & $\epsilon^{-1}(1-10\sigma_2^2+\cdots)$ & $(3\epsilon)^{-1}(1+10\sigma_2^2+\cdots)$
& $1/3$\\ Uniform density & $\sigma_2^{-2}/(3\epsilon)$ & $\sigma_2^{-2}/(3\epsilon)$ & $1$ & $(3\epsilon)^{-1}(\sigma_2^{-2}-13+\cdots)$ & $(3\epsilon)^{-1}(\sigma_2^{-2}-13/3+\cdots)$ &
$1$
\end{tabular}
\end{ruledtabular}
\caption{Gauge-fixed scalar 2EMT in the short-wavelength domain, comparing the strict UV and intermediate UV regimes.
The leading-order terms in the intermediate UV regime coincide with those in the strict UV
limit, so the leading values of $w_{\rm raw}$ are unchanged.
The additional terms displayed in the intermediate-UV columns show the first finite-wavelength gradient corrections, and
make the gauge-dependent gradient differences explicit.}
\label{tab:SWsummary}

\end{table}
\FloatBarrier

\subsection{Strict IR}

In the longitudinal and spatially-flat gauges, the lowest nonvanishing strict-IR terms are slow-roll suppressed,
\begin{equation}
\widehat\tau_{00}^{\lgauge}\simeq\widehat\tau_{00}^{\sfgauge}
\simeq-(9\epsilon^2-3\epsilon\delta)\factLW,
\qquad
\widehat\tau_{ii}^{\lgauge}\simeq\widehat\tau_{ii}^{\sfgauge}
\simeq(9\epsilon^2-3\epsilon\delta)\factLW.
\label{eq:LSFIR}
\end{equation}
The comoving and uniform-density sources start at $\Order(\sigma_1^2)$, with respective leading coefficient pairs $(-10,10/3)$ and
$(-18,6)$ in units of $\sigma_1^2\factLW$. Consequently, their strict-IR ratio limits are $-1/3$ at leading slow-roll
order, with linear corrections $-7\epsilon/15$ and $-11\epsilon/27$, respectively. Their vanishing values at exactly $k=0$ satisfy
the homogeneous constant-$\epsilon$ benchmark; the undefined ratio $0/0$ at that single point does not replace
the ratio limit.

\subsection{Intermediate IR}

The intermediate hierarchy $\epsilon,|\delta|\ll\sigma_1^2\ll1$ changes the dominant balance of the geometrical gauges, as summarized in
Table~\ref{tab:LWsummary}, and the longitudinal and spatially-flat gauges move from the vacuum-like ratio $w_{\rm raw}\simeq-1$ in
the strict IR to the gradient-dominated ratio $w_{\rm raw}\simeq-1/3$ in the intermediate IR.
The comoving and uniform-density sources instead begin at $\Order(\sigma_1^2)$, with different leading coefficients but the
same leading ratio $-1/3$. Notably, all four gauges have $w_{\rm raw}\simeq-1/3$ at leading intermediate-IR order,
although the magnitudes of the individual 2EMT components remain strongly gauge dependent.
Agreement of the ratio therefore does not imply agreement of the 2EMT.

\subsection{Strict UV}

The strict-UV hierarchy is qualitatively distinct among the four gauges, as summarized in Table~\ref{tab:SWsummary}.
The longitudinal and spatially-flat gauges give the radiationlike value $w_{\rm raw}=1/3$.
The comoving gauge is enhanced by $1/\epsilon$, and has positive leading density and pressure, with
$w_{\rm raw}=1/3$.
The uniform-density gauge is more strongly gradient enhanced, by an additional $\sigma_2^{-2}$ relative to the
common $\factSW$ baseline, and has positive density and pressure, with $w_{\rm raw}=1$.

Both are matter-clock gauges, and both inherit a $1/\epsilon$ enhancement because the corresponding background clocks
evolve slowly during inflation.
The uniform-density clock additionally contains the Laplacian term in Eq.~\eqref{eq:Qdifference}, which changes the dominant ultraviolet
structure, and produces the extra factor $\sigma_2^{-2}$.
In particular,
\begingroup
\begin{equation}
\frac{\widehat\tau_{00}^{\ud}}{|\widehat\tau_{00}^{\cgauge}|}
\sim\frac{1}{3\sigma_2^2}\gg1
\end{equation}
\endgroup
for sufficiently short wavelengths at fixed slow roll.

\subsection{Intermediate UV}

For the hierarchy $\epsilon,\delta\ll\sigma_2^2\ll1$, Table~\ref{tab:SWsummary} shows the first finite-wavelength corrections to the strict-UV results.
The leading terms are unchanged from the strict UV limit, and consequently the leading values
of $w_{\rm raw}$ are unchanged as well.
The new information in the intermediate regime is therefore contained in the subleading gradient structure.

For the longitudinal and spatially-flat gauges, the first slow-roll-independent corrections to $\widehat\tau_{00}/\factSW$ are both $\sigma_2^2$,
respectively, while $\widehat\tau_{ii}/\factSW$ receives different finite-wavelength corrections.
The comoving and uniform-density gauges retain their respective $1/\epsilon$ and additional $\sigma_2^{-2}$ enhancements, while the
finite-wavelength terms become explicitly ordered.
Thus no new gauge-family crossover occurs in passing from the strict UV limit to the
intermediate UV regime.
Rather, the intermediate expansion resolves the gauge-dependent finite-wavelength corrections around the same leading UV structures.

\section{Interpretation: physical clocks, gauge fixing, and observability}

\label{sec:interpretation}

The uniform-density gauge has a direct slicing interpretation: coordinate time is chosen so that each
hypersurface has no first-order density perturbation.
Nevertheless, the quantity calculated here remains a raw gauge-fixed 2EMT.
Complete gauge fixing selects one representative of the perturbative diffeomorphism orbit, but does not by
itself construct an {\it observable}.

The comparison with the recalculated longitudinal, spatially-flat, and comoving sectors sharpens the physical-clock interpretation.
The longitudinal and spatially-flat gauges are primarily geometrical slicings, and do not define their time
coordinates through a slowly evolving matter clock.
By contrast, the comoving and uniform-density gauges use the inflaton and density, respectively, as ``matter
clocks''.
Schematically, a clock-defined time shift has the structure
\begin{equation}
\xi_X^0\sim\frac{\delta X}{X_0'}.
\end{equation}
During slow roll, $\phi_0'^2\propto\epsilon$ and $\rho_0'\propto\epsilon$, so the corresponding slicings become increasingly ill conditioned as
the de Sitter limit is approached.
At fixed $c_i$, $a$, and $\cH$, this is the origin of the common $1/\epsilon$ enhancement
in the comoving and uniform-density short-wavelength results.
The density clock carries the additional Laplacian contribution in Eq.~\eqref{eq:Qdifference}, producing the extra $1/\sigma_2^2$ factor
in the uniform-density gauge.

The ``strict/intermediate'' distinction also clarifies which enhancements are intrinsic to a chosen clock and which
simply reflect derivative scaling.
In the ultraviolet, inverse powers of $\sigma_2$ correspond to positive powers of $k/\cH$.
The common $1/\sigma_2^4$ behavior before extracting $\factSW$ has $k^4$ scaling at fixed $c_i$ of a
derivative-dominated quadratic source, not by itself a gauge pathology.
Gauge-specific enhancement is identified only after the common short-wavelength scaling has been factored out.
In the infrared, the ``intermediate'' expansion makes finite-gradient differences visible even when both sources vanish
at zero gradient; their first nonzero coefficients need not agree.

A direct connection between the present gauge-fixed source and observation lies beyond the scope of
this work.
Nevertheless, the gauge comparison suggests a possible direction for future study.
For example, if one wishes to associate the 2EMT with an observer-dependent effective energy density,
a natural starting point is the projection $\tau_{\mu\nu}^{(2)}u^\mu u^\nu$.
This projection is only one illustrative candidate and is not, by itself, a complete observable.
A relational completion would additionally require a physical clock $X$, the same observer congruence $u^\mu$,
the hypersurface $X=X_*$, and a specified measurement or coarse-graining prescription in every gauge.
Schematically, one may consider
\begin{equation}
\mathscr O_{\twoemt}^{G}(X_*)=
\left[
\tau_{\mu\nu,G}^{(2)}u_G^\mu u_G^\nu
+\Delta_{{\rm clock},G}
+\Delta_{{\rm observer},G}
\right]_{X_G=X_*}.
\label{eq:relational}
\end{equation}
The individual terms and their coordinate expressions generally differ among gauges.
Only a fully specified completion, evaluated on the same physical hypersurface with the same physical
observer, would be expected to agree across gauge representations.
The raw quantities $\tau_{\mu\nu,G}^{(2)}$ need not agree, and Eq.~\eqref{eq:relational} is intended only as a possible
future construction rather than a unique definition of an observable.

Within this hierarchy, the present uniform-density result is useful in three ways.
First, it identifies the quadratic source associated with a concrete matter clock.
Second, its vanishing zero-gradient source agrees with the comoving result and with the homogeneous constant-$\epsilon$
benchmark. Their first nonzero coefficients differ, while both strict-IR ratio limits are $-1/3$ at leading
slow-roll order; this is not gauge invariance of the 2EMT.
Third, the strict and intermediate ultraviolet results distinguish the common matter-clock enhancement from the additional
gradient sensitivity of the density clock.
The first three gauges were already considered in Ref.~\cite{Cho:2022maa}, but the present recalculation and three-stage
presentation expose their regime dependence more explicitly and lead to a refined comparison of the
dominant terms.
A companion work will treat the uniform-expansion gauge separately.

\section{Conclusions}

\label{sec:conclusion}

We have calculated the scalar 2EMT on uniform-density hypersurfaces during slow-roll inflation, and compared it
with newly recalculated longitudinal, spatially-flat, and comoving results.
In contrast with an analysis based on a single formal long- or short-wavelength expansion, the
present expressions distinguish the strict IR/UV limits from the intermediate regimes $\epsilon,|\delta|\ll\sigma_i^2\ll1$, and display each
result in full $k$-space, $\sigma_i$, and slow-roll-reordered forms.

First, the uniform-density gauge is fixed by $\delta\rho=E=0$, and its time-shift variable is
\begin{equation}
Q_{\ud}=\beta_{\ud}
=\frac{3\cH(\Psi'+\cH\Psi)-\Delta\Psi}{3\cH(\cH'-\cH^2)}.
\end{equation}
Its difference from the comoving shift is purely gradient dependent, through the $\Delta\Psi$ term.
The uniform-density and comoving 2EMTs have vanishing constant terms in the strict adiabatic IR limit,
although their first nonzero gradient coefficients differ.

Second, in the strict IR,
\begingroup
\begin{equation}
\widehat\tau_{00}^{\ud}\simeq-18\sigma_1^2\factLW,
\qquad
\widehat\tau_{ii}^{\ud}\simeq6\sigma_1^2\factLW,
\qquad
w_{\rm raw}^{\ud}\simeq-\frac13.
\end{equation}
\endgroup
The comoving coefficients are $(-10,10/3)\sigma_1^2\factLW$, where the pair refers to the $(00,ii)$ components, respectively. The
common ratio is not a universal gauge-independent value of the 2EMT.
In the intermediate IR, the sources begin at $\Order(\sigma_1^2)$, and the explicit $\sigma_1^2$, $\sigma_1^4$, and
higher gradient coefficients separate the two gauges.
The longitudinal and spatially-flat gauges, by contrast, are slow-roll suppressed in the strict IR and
become gradient dominated in the intermediate IR, where all four gauges happen to give $w_{\rm
raw}\simeq-1/3$ at leading order, although the magnitudes of the individual 2EMT components remain strongly gauge
dependent.

Third, in the strict UV, the longitudinal, spatially-flat, and comoving ratios are $1/3$, whereas for
uniform density,
\begingroup
\begin{equation}
\widehat\tau_{00}^{\ud}\simeq\frac{\factSW}{3\epsilon\sigma_2^2},
\qquad
\widehat\tau_{ii}^{\ud}\simeq\frac{\factSW}{3\epsilon\sigma_2^2},
\qquad
w_{\rm raw}^{\ud}\simeq1.
\end{equation}
\endgroup
The intermediate UV expansion preserves this leading structure while explicitly ordering the subleading gradient corrections.
The common $1/\epsilon$ enhancement of the comoving and uniform-density gauges reflects the slow evolution of
their respective matter clocks.
The uniform-density result contains the additional factor $1/\sigma_2^2$ because the density constraint includes a Laplacian.
These factors are properties of the gauge clocks and should not be read as gauge-independent
physical instabilities.

Finally, the four-gauge comparison reveals structured rather than arbitrary dependence.
The strict IR separates the geometrical pair, longitudinal and spatially flat, from the matter-clock pair,
comoving and uniform density.
The intermediate IR shows that the component ratio alone can become common while the magnitudes
of the individual 2EMT components remain gauge dependent.
In the ultraviolet, the common inverse powers associated with $k^4$ derivative scaling must be separated
from genuinely gauge-specific clock enhancements.
This regime-by-regime structure provides a more precise basis for constructing a future relational completion of
the 2EMT, and for distinguishing clock-dependent effects from quantities that may survive in an observable
prescription.

\begin{acknowledgments}
The author is grateful to Jaichan Hwang for very helpful discussions.
The author used ChatGPT (OpenAI) as an assistive tool during the analysis and preparation of
this manuscript.
All AI-assisted content was critically evaluated and verified by the author, who takes full responsibility
for the scientific content and conclusions.
\end{acknowledgments}

\appendix

\section{General Fourier-space expression for the 2EMT}

\label{app:general2EMT}

For reference, we define the normalized Fourier-space contribution associated with a fixed mode pair $\bm
k$ and $-\bm k$ by
\begin{equation}
\widehat\tau_{\mu\nu}(k)\equiv8\pi G\tau_{\mu\nu}(k)
=8\pi G\left\langle T^{(2,\mathrm{eff})}_{\mu\nu}\right\rangle_k.
\label{eq:hattauDef}
\end{equation}
Before imposing a gauge condition, its scalar components are \cite{Cho:2022maa}
\begin{align}
\widehat\tau_{00}(k)={}&\frac{1}{\cH'-\cH^2}\Bigg\{
-\langle(\nabla\Psi')^2\rangle
-\langle(\Delta\Psi)^2\rangle
-2(K+\cH)\langle\nabla\Psi'\cdot\nabla\Psi\rangle
\nonumber\\
&-\left[3(\cH'-\cH^2)+K^2+a^2V_{\phi\phi}\right]
\langle(\Psi')^2\rangle
+(9\cH'-10\cH^2-2L)\langle(\nabla\Psi)^2\rangle
\nonumber\\
&+2\left[6\cH(\cH'-\cH^2)-KL
+2(\cH'-\cH^2)a^2\frac{V_\phi}{\phi_0'}
-a^2\cH V_{\phi\phi}\right]
\langle\Psi\Psi'\rangle
\nonumber\\
&-\left[L^2-4\cH(\cH'-\cH^2)a^2\frac{V_\phi}{\phi_0'}
+a^2\cH^2V_{\phi\phi}\right]
\langle\Psi^2\rangle\Bigg\}
\nonumber\\
&+2\langle\left\{
a^2\frac{V_\phi}{\phi_0'}Q'
+\left[3\cH'+(4\cH+K)a^2\frac{V_\phi}{\phi_0'}
+a^2V_{\phi\phi}\right]Q
-\Delta Q+\Delta E'-2\cH\Delta E\right\}\Psi'\rangle
\nonumber\\
&+2\langle\Bigg\{
-\left[L+6\cH^2-2a^2\cH\frac{V_\phi}{\phi_0'}\right]Q'
+\left[2\cH(L-3\cH')
+(L-2\cH'+4\cH^2)a^2\frac{V_\phi}{\phi_0'}
+\cH a^2V_{\phi\phi}\right]Q
\nonumber\\
&\qquad-\Delta Q'-(K-\cH)\Delta Q
+2\cH\Delta E'+\Delta^2E\Bigg\}\Psi\rangle
-(\cH'+2\cH^2)\langle(Q')^2\rangle
\nonumber\\
&-2\left[\cH(\cH'-4\cH^2)
+(\cH'-\cH^2)a^2\frac{V_\phi}{\phi_0'}\right]
\langle Q'Q\rangle
\nonumber\\
&-\Bigg\{(\cH'-\cH^2)\left[4\cH^2
+8\cH a^2\frac{V_\phi}{\phi_0'}
+a^4\left(\frac{V_\phi}{\phi_0'}\right)^2
+a^2V_{\phi\phi}\right]
+3\cH'(\cH'-4\cH^2)\Bigg\}
\langle Q^2\rangle
\nonumber\\
&-2\cH\langle\nabla Q'\cdot\nabla Q\rangle
+\langle\left[4\cH^2Q+2\cH\Delta E
-(\cH'+2\cH^2)E'\right]\Delta E'\rangle
+4\cH\langle(\cH Q'+\cH'Q)\Delta E\rangle,
\label{eq:generalAvg00}
\end{align}
After the isotropic projection,
\begingroup
\begin{equation}
\widehat\tau_{0i}=0,\qquad
\widehat\tau_{ij}=\delta_{ij}\widehat\tau_{ii}.
\label{eq:generalAvgOffdiag}
\end{equation}
\endgroup
\begin{align}
\widehat\tau_{ii}(k)={}&\frac{1}{\cH'-\cH^2}\Bigg[
\frac13\langle(\nabla\Psi')^2\rangle
-\langle(\Delta\Psi)^2\rangle
+2(\tfrac{\cH}{3}-K)
\langle\nabla\Psi'\cdot\nabla\Psi\rangle
\nonumber\\
&+(\cH'-\cH^2-K^2+a^2V_{\phi\phi})
\langle(\Psi')^2\rangle
+(\tfrac{11}{3}\cH'-\tfrac{10}{3}\cH^2-2L)
\langle(\nabla\Psi)^2\rangle
\nonumber\\
&+2\left(\cH a^2V_{\phi\phi}-KL
+2(\cH'-\cH^2)(K+2\cH)\right)
\langle\Psi\Psi'\rangle
\nonumber\\
&+\left(4(\cH'-\cH^2)(\cH'+2\cH^2+L)-L^2
+\cH^2a^2V_{\phi\phi}\right)
\langle\Psi^2\rangle\Bigg]
\nonumber\\
&+2\langle\left[Q''+3(3\cH-K)Q'
+(\cH'+12\cH^2+K^2-7\cH K-a^2V_{\phi\phi})Q
+\frac13\Delta E'-\frac43(K-\cH)\Delta E\right]\Psi'\rangle
\nonumber\\
&+2\langle\Bigg[4\cH Q''+(6\cH'+10\cH^2-3L)Q'
+\Delta Q'-(K-3\cH)\Delta Q
\nonumber\\
&\qquad-\left(2\cH''+a^2\cH V_{\phi\phi}-\cH K^2
+(5\cH'+2\cH^2)K-11\cH'\cH-9\cH^3\right)Q
\nonumber\\
&\qquad-\frac13\Delta E''-\frac23\cH\Delta E'
-\frac43L\Delta E+\frac23\Delta^2E\Bigg]\Psi\rangle
\nonumber\\
&+\frac23\langle\nabla Q''\cdot\nabla Q\rangle
+\frac23\langle(\nabla Q')^2\rangle
+\frac{4\cH}{3}\langle\nabla Q'\cdot\nabla Q\rangle
+2\cH\langle Q''Q'\rangle
-2\cH'\langle Q''Q\rangle
-(\cH'-2\cH^2)\langle(Q')^2\rangle
\nonumber\\
&-2\left[2\cH''+3\cH'\cH
+(\cH'-\cH^2)(K+\cH)\right]
\langle Q'Q\rangle
\nonumber\\
&-\Bigg\{4\cH\cH''+3(\cH')^2+4\cH'\cH^2
+(\cH'-\cH^2)\Bigg[4\cH^2-a^2V_{\phi\phi}
+a^4\left(\frac{V_\phi}{\phi_0'}\right)^2\Bigg]\Bigg\}
\langle Q^2\rangle
\nonumber\\
&-\frac23\langle(2\cH Q-3\cH E'+\Delta E)\Delta E''\rangle
\nonumber\\
&-\frac13\langle\left[8\cH Q'+8(\cH'+\cH^2)Q
-3(\cH'+2\cH^2)E'+2\Delta E'+4\cH\Delta E\right]\Delta E'\rangle
\nonumber\\
&-\frac43\left\langle
[\cH Q''+2(\cH'+\cH^2)Q'
+2(3\cH\cH'-\cH^3-(\cH'-\cH^2)K)Q]\Delta E
\right\rangle.
\label{eq:generalAvgii}
\end{align}

In Eqs.~\eqref{eq:generalAvg00}--\eqref{eq:generalAvgii}, $\Psi$, $Q$, and $E$ are expanded in Fourier modes.
The brackets denote the Fourier-transformed quadratic quantities obtained through the required spatial integration; they do
not represent a separate averaging prescription.
For the fixed contribution associated with the pair $\bm k$ and $-\bm k$, the common
momentum labels on $\Psi$, $Q$, and $E$ are suppressed for notational simplicity.
For real perturbations, the $-\bm k$ mode is the complex conjugate of the $\bm k$
mode.
Pairing the modes $\bm k$ and $-\bm k$ does not remove their directional dependence.
The isotropic angular projection is applied separately.

\section{Longitudinal gauge}

\label{app:L}

The longitudinal gauge is defined by $\beta=E=0$, so $Q=0$.
Although the longitudinal, spatially-flat, comoving gauges were analyzed in Ref.~\cite{Cho:2022maa}, we recalculate them in Appendix
B--D directly from the general Fourier-space 2EMT and display the results more explicitly than in
the earlier work.

\subsection{Long-wavelength expressions}
\begingroup
\begin{align}
\widehat\tau_{00}^{\mathrm{LW}}={}&\frac{|A_1|^2 \epsilon}{\cH^{2}}\Big\{k^{4}
+k^{2}\cH^{2}[1+(11 \epsilon-2 \delta)+(4 \epsilon^{2}-4 \delta \epsilon)]
+\cH^{4}[(-9 \epsilon+3 \delta)\nonumber\\
&\hspace{18mm}
+(13 \epsilon^{2}
+2 \delta \epsilon-11 \delta^{2}-\xi_3)+(16 \delta \epsilon^{2}-24 \epsilon \delta^{2}-4 \epsilon \xi_3+8 \delta^{3}
+4 \delta \xi_3)\nonumber\\
&\hspace{18mm}
+(4 \delta^{2} \epsilon^{2}-4 \xi_3 \epsilon^{2}-8 \epsilon \delta^{3}
+8 \delta \epsilon \xi_3+4 \delta^{4}-4 \xi_3 \delta^{2})]\Big\}
\nonumber\\
={}&|A_1|^2 \epsilon \cH^{2}\Big\{[(-9 \epsilon+3 \delta)+(13 \epsilon^{2}
+2 \delta \epsilon-11 \delta^{2}-\xi_3)+(16 \delta \epsilon^{2}-24 \epsilon \delta^{2}-4 \epsilon \xi_3+8 \delta^{3}
+4 \delta \xi_3)\nonumber\\
&\hspace{18mm}
+(4 \delta^{2} \epsilon^{2}-4 \xi_3 \epsilon^{2}-8 \epsilon \delta^{3}
+8 \delta \epsilon \xi_3+4 \delta^{4}-4 \xi_3 \delta^{2})]
+[1+(11 \epsilon-2 \delta)+(4 \epsilon^{2}-4 \delta \epsilon)]\sigma_1^{2}\nonumber\\
&\hspace{18mm}
+\sigma_1^{4}\Big\}\;\text{(strict IR)}
\nonumber\\
\simeq{}&|A_1|^2 \epsilon \cH^{2}\Big\{[\sigma_1^{2}
+\sigma_1^{4}]\nonumber\\
&\hspace{18mm}
+[(-9 \epsilon+3 \delta)+(11 \epsilon-2 \delta)\sigma_1^{2}]
+\Order_{\rm SR}(2;\sigma_1)\Big\}\;\text{(intermediate IR)}.
\label{eq:Llw00}
\end{align}
\endgroup
\begingroup
\begin{align}
\widehat\tau_{ii}^{\mathrm{LW}}={}&\frac{|A_1|^2 \epsilon}{\cH^{2}}\Big\{k^{4}
+k^{2}\cH^{2}[-\frac{1}{3}
+(\frac{\epsilon}{3}
+\frac{10 \delta}{3})+(-\frac{4 \epsilon^{2}}{3}
+\frac{20 \delta \epsilon}{3}-\frac{16 \delta^{2}}{3})]
+\cH^{4}[(9 \epsilon-3 \delta)\nonumber\\
&\hspace{18mm}
+(-3 \epsilon^{2}-14 \delta \epsilon+13 \delta^{2}
+\xi_3)+(-8 \epsilon^{3}
+8 \delta \epsilon^{2}
+16 \epsilon \delta^{2}
+4 \epsilon \xi_3-16 \delta^{3}-4 \delta \xi_3)\nonumber\\
&\hspace{18mm}
+(4 \delta^{2} \epsilon^{2}
+4 \xi_3 \epsilon^{2}-8 \epsilon \delta^{3}-8 \delta \epsilon \xi_3+4 \delta^{4}
+4 \xi_3 \delta^{2})]\Big\}
\nonumber\\
={}&|A_1|^2 \epsilon \cH^{2}\Big\{[(9 \epsilon-3 \delta)+(-3 \epsilon^{2}-14 \delta \epsilon+13 \delta^{2}
+\xi_3)+(-8 \epsilon^{3}
+8 \delta \epsilon^{2}
+16 \epsilon \delta^{2}
+4 \epsilon \xi_3-16 \delta^{3}-4 \delta \xi_3)\nonumber\\
&\hspace{18mm}
+(4 \delta^{2} \epsilon^{2}
+4 \xi_3 \epsilon^{2}-8 \epsilon \delta^{3}-8 \delta \epsilon \xi_3+4 \delta^{4}
+4 \xi_3 \delta^{2})]\nonumber\\
&\hspace{18mm}
+[-\frac{1}{3}
+(\frac{\epsilon}{3}
+\frac{10 \delta}{3})+(-\frac{4 \epsilon^{2}}{3}
+\frac{20 \delta \epsilon}{3}-\frac{16 \delta^{2}}{3})]\sigma_1^{2}
+\sigma_1^{4}\Big\}\;\text{(strict IR)}
\nonumber\\
\simeq{}&|A_1|^2 \epsilon \cH^{2}\Big\{[-\frac{1}{3}\sigma_1^{2}
+\sigma_1^{4}]\nonumber\\
&\hspace{18mm}
+[(9 \epsilon-3 \delta)+(\frac{\epsilon}{3}
+\frac{10 \delta}{3})\sigma_1^{2}]
+\Order_{\rm SR}(2;\sigma_1)\Big\}\;\text{(intermediate IR)}.
\label{eq:Llwii}
\end{align}
\endgroup
The strict-IR leading terms are $(-3\epsilon(3\epsilon-\delta),3\epsilon(3\epsilon-\delta))\factLW$.
In the intermediate hierarchy the gradient terms instead dominate, giving $(\epsilon,-\epsilon/3)\sigma_1^2\factLW$ and $w_{\rm raw}\simeq-1/3$. \subsection{Short-wavelength
expressions}
\begingroup
\begin{align}
\widehat\tau_{00}^{\mathrm{SW}}={}&\frac{2 \pi C_k G}{a^{2}}\Big\{2k^{4}
+k^{2}\cH^{2}[1+(7 \epsilon+3 \delta)+(-\xi_3)]
+\cH^{4}[(-9 \epsilon+3 \delta)+(\epsilon^{2}-4 \delta \epsilon-5 \delta^{2}-\xi_3)\nonumber\\
&\hspace{18mm}
+(-2 \epsilon \delta^{2}
+\delta^{3}
+2 \delta \xi_3)+(\delta^{4}-\xi_3 \delta^{2})]\Big\}
\nonumber\\
={}&\frac{2 \pi C_k G \cH^{4}}{a^{2}}\Big\{2\sigma_2^{-4}
+[1+(7 \epsilon+3 \delta)+(-\xi_3)]\sigma_2^{-2}
+[(-9 \epsilon+3 \delta)+(\epsilon^{2}-4 \delta \epsilon-5 \delta^{2}-\xi_3)\nonumber\\
&\hspace{18mm}
+(-2 \epsilon \delta^{2}
+\delta^{3}
+2 \delta \xi_3)+(\delta^{4}-\xi_3 \delta^{2})]\Big\}\;\text{(strict UV)}
\nonumber\\
\simeq{}&\frac{2 \pi C_k G \cH^{4}}{a^{2}}\Big\{[2\sigma_2^{-4}
+\sigma_2^{-2}]\nonumber\\
&\hspace{18mm}
+[(7 \epsilon+3 \delta)\sigma_2^{-2}
+(-9 \epsilon+3 \delta)]
+\Order_{\rm SR}(2;\sigma_2)\Big\}\;\text{(intermediate UV)}.
\label{eq:Lsw00}
\end{align}
\endgroup
\begingroup
\begin{align}
\widehat\tau_{ii}^{\mathrm{SW}}={}&\frac{2 \pi C_k G}{3 a^{2}}\Big\{2k^{4}
+k^{2}\cH^{2}[-1+(-\epsilon-\delta)+(-4 \delta^{2}
+3 \xi_3)]
+\cH^{4}[(27 \epsilon-9 \delta)+(-21 \epsilon^{2}
+21 \delta^{2}
+3 \xi_3)\nonumber\\
&\hspace{18mm}
+(-12 \epsilon \delta^{2}-15 \delta^{3}-6 \delta \xi_3)+(3 \delta^{4}
+3 \xi_3 \delta^{2})]\Big\}
\nonumber\\
={}&\frac{2 \pi C_k G \cH^{4}}{3 a^{2}}\Big\{2\sigma_2^{-4}
+[-1+(-\epsilon-\delta)+(-4 \delta^{2}
+3 \xi_3)]\sigma_2^{-2}
+[(27 \epsilon-9 \delta)+(-21 \epsilon^{2}
+21 \delta^{2}
+3 \xi_3)\nonumber\\
&\hspace{18mm}
+(-12 \epsilon \delta^{2}-15 \delta^{3}-6 \delta \xi_3)+(3 \delta^{4}
+3 \xi_3 \delta^{2})]\Big\}\;\text{(strict UV)}
\nonumber\\
\simeq{}&\frac{2 \pi C_k G \cH^{4}}{3 a^{2}}\Big\{[2\sigma_2^{-4}-\sigma_2^{-2}]\nonumber\\
&\hspace{18mm}
+[(-\epsilon-\delta)\sigma_2^{-2}
+(27 \epsilon-9 \delta)]
+\Order_{\rm SR}(2;\sigma_2)\Big\}\;\text{(intermediate UV)}.
\label{eq:Lswii}
\end{align}
\endgroup
At highest power in $k/\cH$, the same result is visible directly from the general source,
\begingroup
\begin{equation}
\widehat\tau_{00}^{\lgauge}
\simeq\frac{-k^2|\Psi_k'|^2-k^4|\Psi_k|^2}{\cH'-\cH^2},
\qquad
\widehat\tau_{ii}^{\lgauge}
\simeq\frac{k^2|\Psi_k'|^2/3-k^4|\Psi_k|^2}{\cH'-\cH^2}.
\end{equation}
\endgroup
With $|\Psi_k'|^2\simeq k^2|\Psi_k|^2$ and $\cH'-\cH^2=-\epsilon\cH^2$, this gives $\widehat\tau_{00}/\factSW\to2$ and $\widehat\tau_{ii}/\factSW\to2/3$.

\section{Spatially-flat gauge}

\label{app:SF}

The spatially-flat gauge is defined by $\psi=E=0$, implying $Q=\Psi/\cH$.

\subsection{Long-wavelength expressions}
\begingroup
\begin{align}
\widehat\tau_{00}^{\mathrm{LW}}={}&\frac{|A_1|^2 \epsilon}{\cH^{2}}\Big\{k^{4}
+k^{2}\cH^{2}[1+(9 \epsilon-2 \delta)+(8 \epsilon^{2}-6 \delta \epsilon)]
+\cH^{4}[(-9 \epsilon+3 \delta)\nonumber\\
&\hspace{18mm}
+(-53 \epsilon^{2}
+56 \delta \epsilon-11 \delta^{2}-\xi_3)+(-75 \epsilon^{3}
+143 \delta \epsilon^{2}-74 \epsilon \delta^{2}-6 \epsilon \xi_3+8 \delta^{3}
+4 \delta \xi_3)\nonumber\\
&\hspace{18mm}
+(9 \epsilon^{4}
+6 \delta \epsilon^{3}-11 \delta^{2} \epsilon^{2}-9 \xi_3 \epsilon^{2}-4 \epsilon \delta^{3}
+12 \delta \epsilon \xi_3+4 \delta^{4}-4 \xi_3 \delta^{2})]\Big\}
\nonumber\\
={}&|A_1|^2 \epsilon \cH^{2}\Big\{[(-9 \epsilon+3 \delta)+(-53 \epsilon^{2}
+56 \delta \epsilon-11 \delta^{2}-\xi_3)\nonumber\\
&\hspace{18mm}
+(-75 \epsilon^{3}
+143 \delta \epsilon^{2}-74 \epsilon \delta^{2}-6 \epsilon \xi_3+8 \delta^{3}
+4 \delta \xi_3)\nonumber\\
&\hspace{18mm}
+(9 \epsilon^{4}
+6 \delta \epsilon^{3}-11 \delta^{2} \epsilon^{2}-9 \xi_3 \epsilon^{2}-4 \epsilon \delta^{3}
+12 \delta \epsilon \xi_3+4 \delta^{4}-4 \xi_3 \delta^{2})]\nonumber\\
&\hspace{18mm}
+[1+(9 \epsilon-2 \delta)+(8 \epsilon^{2}-6 \delta \epsilon)]\sigma_1^{2}
+\sigma_1^{4}\Big\}\;\text{(strict IR)}
\nonumber\\
\simeq{}&|A_1|^2 \epsilon \cH^{2}\Big\{[\sigma_1^{2}
+\sigma_1^{4}]\nonumber\\
&\hspace{18mm}
+[(-9 \epsilon+3 \delta)+(9 \epsilon-2 \delta)\sigma_1^{2}]
+\Order_{\rm SR}(2;\sigma_1)\Big\}\;\text{(intermediate IR)}.
\label{eq:SFlw00}
\end{align}
\endgroup
\begingroup
\begin{align}
\widehat\tau_{ii}^{\mathrm{LW}}={}&\frac{|A_1|^2 \epsilon}{3 \cH^{2}}\Big\{k^{4}[3+(-2 \epsilon)]
+k^{2}\cH^{2}[-1+(-25 \epsilon+10 \delta)+(-62 \epsilon^{2}
+66 \delta \epsilon-16 \delta^{2})\nonumber\\
&\hspace{18mm}
+(30 \epsilon^{3}-44 \delta \epsilon^{2}
+16 \epsilon \delta^{2})]
+\cH^{4}[(27 \epsilon-9 \delta)+(189 \epsilon^{2}-204 \delta \epsilon+39 \delta^{2}
+3 \xi_3)\nonumber\\
&\hspace{18mm}
+(405 \epsilon^{3}-765 \delta \epsilon^{2}
+402 \epsilon \delta^{2}
+18 \epsilon \xi_3-48 \delta^{3}-12 \delta \xi_3)\nonumber\\
&\hspace{18mm}
+(243 \epsilon^{4}-702 \delta \epsilon^{3}
+639 \delta^{2} \epsilon^{2}
+27 \xi_3 \epsilon^{2}-204 \epsilon \delta^{3}-36 \delta \epsilon \xi_3+12 \delta^{4}
+12 \xi_3 \delta^{2})]\Big\}
\nonumber\\
={}&|A_1|^2 \epsilon \cH^{2}\Big\{[(9 \epsilon-3 \delta)+(63 \epsilon^{2}-68 \delta \epsilon+13 \delta^{2}
+\xi_3)\nonumber\\
&\hspace{18mm}
+(135 \epsilon^{3}-255 \delta \epsilon^{2}
+134 \epsilon \delta^{2}
+6 \epsilon \xi_3-16 \delta^{3}-4 \delta \xi_3)\nonumber\\
&\hspace{18mm}
+(81 \epsilon^{4}-234 \delta \epsilon^{3}
+213 \delta^{2} \epsilon^{2}
+9 \xi_3 \epsilon^{2}-68 \epsilon \delta^{3}-12 \delta \epsilon \xi_3+4 \delta^{4}
+4 \xi_3 \delta^{2})]
+[-\frac{1}{3}
+(-\frac{25 \epsilon}{3}
+\frac{10 \delta}{3})\nonumber\\
&\hspace{18mm}
+(-\frac{62 \epsilon^{2}}{3}
+22 \delta \epsilon-\frac{16 \delta^{2}}{3})+(10 \epsilon^{3}-\frac{44 \delta \epsilon^{2}}{3}
+\frac{16 \epsilon \delta^{2}}{3})]\sigma_1^{2}
+[1+(-\frac{2 \epsilon}{3})]\sigma_1^{4}\Big\}\;\text{(strict IR)}
\nonumber\\
\simeq{}&|A_1|^2 \epsilon \cH^{2}\Big\{[-\frac{1}{3}\sigma_1^{2}
+\sigma_1^{4}]\nonumber\\
&\hspace{18mm}
+[(9 \epsilon-3 \delta)+(-\frac{25 \epsilon}{3}
+\frac{10 \delta}{3})\sigma_1^{2}-\frac{2 \epsilon}{3}\sigma_1^{4}]
+\Order_{\rm SR}(2;\sigma_1)\Big\}\;\text{(intermediate IR)}.
\label{eq:SFlwii}
\end{align}
\endgroup
The strict-IR leading terms are $(-3\epsilon(3\epsilon-\delta),3\epsilon(3\epsilon-\delta))\factLW$.
In the intermediate hierarchy the gradient terms instead dominate, giving $(\epsilon,-\epsilon/3)\sigma_1^2\factLW$ and $w_{\rm raw}\simeq-1/3$. \subsection{Short-wavelength
expressions}
\begingroup
\begin{align}
\widehat\tau_{00}^{\mathrm{SW}}={}&\frac{2 \pi C_k G}{a^{2}}\Big\{2k^{4}
+k^{2}\cH^{2}[1+(-4 \epsilon+3 \delta)+(\epsilon^{2}
+2 \delta \epsilon-\xi_3)]
+\cH^{4}[(-9 \epsilon+3 \delta)\nonumber\\
&\hspace{18mm}
+(-17 \epsilon^{2}
+26 \delta \epsilon-5 \delta^{2}-\xi_3)+(-7 \epsilon^{3}
+23 \delta \epsilon^{2}-17 \epsilon \delta^{2}-2 \epsilon \xi_3+\delta^{3}
+2 \delta \xi_3)\nonumber\\
&\hspace{18mm}
+(\epsilon^{4}-2 \delta^{2} \epsilon^{2}-\xi_3 \epsilon^{2}
+2 \delta \epsilon \xi_3+\delta^{4}-\xi_3 \delta^{2})]\Big\}
\nonumber\\
={}&\frac{2 \pi C_k G \cH^{4}}{a^{2}}\Big\{2\sigma_2^{-4}
+[1+(-4 \epsilon+3 \delta)+(\epsilon^{2}
+2 \delta \epsilon-\xi_3)]\sigma_2^{-2}
+[(-9 \epsilon+3 \delta)+(-17 \epsilon^{2}
+26 \delta \epsilon-5 \delta^{2}-\xi_3)\nonumber\\
&\hspace{18mm}
+(-7 \epsilon^{3}
+23 \delta \epsilon^{2}-17 \epsilon \delta^{2}-2 \epsilon \xi_3+\delta^{3}
+2 \delta \xi_3)\nonumber\\
&\hspace{18mm}
+(\epsilon^{4}-2 \delta^{2} \epsilon^{2}-\xi_3 \epsilon^{2}
+2 \delta \epsilon \xi_3+\delta^{4}-\xi_3 \delta^{2})]\Big\}\;\text{(strict UV)}
\nonumber\\
\simeq{}&\frac{2 \pi C_k G \cH^{4}}{a^{2}}\Big\{[2\sigma_2^{-4}
+\sigma_2^{-2}]\nonumber\\
&\hspace{18mm}
+[(-4 \epsilon+3 \delta)\sigma_2^{-2}
+(-9 \epsilon+3 \delta)]
+\Order_{\rm SR}(2;\sigma_2)\Big\}\;\text{(intermediate UV)}.
\label{eq:SFsw00}
\end{align}
\endgroup
\begingroup
\begin{align}
\widehat\tau_{ii}^{\mathrm{SW}}={}&\frac{2 \pi C_k G}{3 a^{2}}\Big\{2k^{4}
+k^{2}\cH^{2}[-1+(6 \epsilon-\delta)+(13 \epsilon^{2}-18 \delta \epsilon-4 \delta^{2}
+3 \xi_3)+(6 \epsilon^{3}-12 \delta \epsilon^{2}
+6 \epsilon \delta^{2})]\nonumber\\
&\hspace{18mm}
+\cH^{4}[(27 \epsilon-9 \delta)+(81 \epsilon^{2}-114 \delta \epsilon+21 \delta^{2}
+3 \xi_3)+(81 \epsilon^{3}-201 \delta \epsilon^{2}
+135 \epsilon \delta^{2}
+6 \epsilon \xi_3-15 \delta^{3}-6 \delta \xi_3)\nonumber\\
&\hspace{18mm}
+(27 \epsilon^{4}-96 \delta \epsilon^{3}
+114 \delta^{2} \epsilon^{2}
+3 \xi_3 \epsilon^{2}-48 \epsilon \delta^{3}-6 \delta \epsilon \xi_3+3 \delta^{4}
+3 \xi_3 \delta^{2})]\Big\}
\nonumber\\
={}&\frac{2 \pi C_k G \cH^{4}}{3 a^{2}}\Big\{2\sigma_2^{-4}
+[-1+(6 \epsilon-\delta)+(13 \epsilon^{2}-18 \delta \epsilon-4 \delta^{2}
+3 \xi_3)+(6 \epsilon^{3}-12 \delta \epsilon^{2}
+6 \epsilon \delta^{2})]\sigma_2^{-2}\nonumber\\
&\hspace{18mm}
+[(27 \epsilon-9 \delta)+(81 \epsilon^{2}-114 \delta \epsilon+21 \delta^{2}
+3 \xi_3)+(81 \epsilon^{3}-201 \delta \epsilon^{2}
+135 \epsilon \delta^{2}
+6 \epsilon \xi_3-15 \delta^{3}-6 \delta \xi_3)\nonumber\\
&\hspace{18mm}
+(27 \epsilon^{4}-96 \delta \epsilon^{3}
+114 \delta^{2} \epsilon^{2}
+3 \xi_3 \epsilon^{2}-48 \epsilon \delta^{3}-6 \delta \epsilon \xi_3+3 \delta^{4}
+3 \xi_3 \delta^{2})]\Big\}\;\text{(strict UV)}
\nonumber\\
\simeq{}&\frac{2 \pi C_k G \cH^{4}}{3 a^{2}}\Big\{[2\sigma_2^{-4}-\sigma_2^{-2}]\nonumber\\
&\hspace{18mm}
+[(6 \epsilon-\delta)\sigma_2^{-2}
+(27 \epsilon-9 \delta)]
+\Order_{\rm SR}(2;\sigma_2)\Big\}\;\text{(intermediate UV)}.
\label{eq:SFswii}
\end{align}
\endgroup
The strict UV limit is $(2,2/3)\factSW$, and $w_{\rm raw}\simeq1/3$.

\section{Comoving gauge}

\label{app:C}

The comoving gauge is defined by $\delta\phi=E=0$, for which $Q=(\Psi'+\cH\Psi)/(\cH'-\cH^2)$.

\subsection{Long-wavelength expressions}
\begingroup
\begin{align}
\widehat\tau_{00}^{\mathrm{LW}}={}&\frac{|A_1|^2}{\cH^{2}}\Big\{k^{4}[-1+(8 \epsilon-4 \delta)]
+k^{2}\cH^{2}[-10+(-25 \epsilon+16 \delta)+(17 \epsilon^{2}-20 \delta \epsilon+8 \delta^{2})\nonumber\\
&\hspace{18mm}
+(-4 \epsilon^{3}
+8 \delta \epsilon^{2}-4 \epsilon \delta^{2})]\Big\}
\nonumber\\
={}&|A_1|^2 \cH^{2}\Big\{[-10+(-25 \epsilon+16 \delta)+(17 \epsilon^{2}-20 \delta \epsilon+8 \delta^{2})+(-4 \epsilon^{3}
+8 \delta \epsilon^{2}-4 \epsilon \delta^{2})]\sigma_1^{2}\nonumber\\
&\hspace{18mm}
+[-1+(8 \epsilon-4 \delta)]\sigma_1^{4}\Big\}\;\text{(strict IR)}
\nonumber\\
\simeq{}&|A_1|^2 \cH^{2}\Big\{[-10\sigma_1^{2}-\sigma_1^{4}]\nonumber\\
&\hspace{18mm}
+[(-25 \epsilon+16 \delta)\sigma_1^{2}
+(8 \epsilon-4 \delta)\sigma_1^{4}]
+\Order_{\rm SR}(2;\sigma_1)\Big\}\;\text{(intermediate IR)}.
\label{eq:Clw00}
\end{align}
\endgroup
\begingroup
\begin{align}
\widehat\tau_{ii}^{\mathrm{LW}}={}&\frac{|A_1|^2}{3 \cH^{4}}\Big\{2k^{6}
+k^{4}\cH^{2}[-9+(16 \epsilon-12 \delta)+(-8 \epsilon^{2}
+8 \delta \epsilon)]
+k^{2}\cH^{4}[10+(39 \epsilon-16 \delta)\nonumber\\
&\hspace{18mm}
+(33 \epsilon^{2}-12 \delta \epsilon-8 \delta^{2})+(12 \epsilon^{3}-24 \delta \epsilon^{2}
+12 \epsilon \delta^{2})]\Big\}
\nonumber\\
={}&|A_1|^2 \cH^{2}\Big\{[\frac{10}{3}
+(13 \epsilon-\frac{16 \delta}{3})+(11 \epsilon^{2}-4 \delta \epsilon-\frac{8 \delta^{2}}{3})+(4 \epsilon^{3}-8 \delta \epsilon^{2}
+4 \epsilon \delta^{2})]\sigma_1^{2}\nonumber\\
&\hspace{18mm}
+[-3+(\frac{16 \epsilon}{3}-4 \delta)+(-\frac{8 \epsilon^{2}}{3}
+\frac{8 \delta \epsilon}{3})]\sigma_1^{4}
+\frac{2}{3}\sigma_1^{6}\Big\}\;\text{(strict IR)}
\nonumber\\
\simeq{}&|A_1|^2 \cH^{2}\Big\{[\frac{10}{3}\sigma_1^{2}-3\sigma_1^{4}
+\frac{2}{3}\sigma_1^{6}]\nonumber\\
&\hspace{18mm}
+[(13 \epsilon-\frac{16 \delta}{3})\sigma_1^{2}
+(\frac{16 \epsilon}{3}-4 \delta)\sigma_1^{4}]
+\Order_{\rm SR}(2;\sigma_1)\Big\}\;\text{(intermediate IR)}.
\label{eq:Clwii}
\end{align}
\endgroup
The first nonzero terms at fixed slow-roll parameters are
\begingroup
\begin{align}
\frac{\widehat\tau_{00}^{\cgauge}}{\factLW}&=-\sigma_1^2\Big\{10+(25\epsilon-16\delta)+\Order_{\rm SR}(2)\Big\}
+\Order(\sigma_1^4),\nonumber\\*
\frac{\widehat\tau_{ii}^{\cgauge}}{\factLW}&=\frac{\sigma_1^2}{3}\Big\{10+(39\epsilon-16\delta)+\Order_{\rm SR}(2)\Big\}
+\Order(\sigma_1^4),\nonumber\\*
\lim_{\sigma_1\to0}w_{\rm raw}^{\cgauge}&=-\frac13-\frac{7\epsilon}{15}
+\Order_{\rm SR}(2).
\label{eq:CstrictIRratio}
\end{align}
\endgroup
Both components vanish at exactly $k=0$, while their ratio has the finite strict-IR limit above.
In the intermediate hierarchy the leading coefficients are likewise $(-10,10/3)\sigma_1^2\factLW$, giving $w_{\rm raw}\simeq-1/3$. \subsection{Short-wavelength expressions}
\begingroup
\begin{align}
\widehat\tau_{00}^{\mathrm{SW}}={}&\frac{2 \pi C_k G}{a^{2} \epsilon}\Big\{k^{4}[1+(3 \epsilon-2 \delta)]
+k^{2}\cH^{2}[-10+(-9 \epsilon+8 \delta)+(5 \epsilon^{2}-2 \delta \epsilon+2 \delta^{2})+(-\epsilon \delta^{2})]\Big\}
\nonumber\\
={}&\frac{2 \pi C_k G \cH^{4}}{a^{2} \epsilon}\Big\{[1+(3 \epsilon-2 \delta)]\sigma_2^{-4}
+[-10+(-9 \epsilon+8 \delta)+(5 \epsilon^{2}-2 \delta \epsilon+2 \delta^{2})\nonumber\\
&\hspace{18mm}
+(-\epsilon \delta^{2})]\sigma_2^{-2}\Big\}\;\text{(strict UV)}
\nonumber\\
\simeq{}&\frac{2 \pi C_k G \cH^{4}}{a^{2} \epsilon}\Big\{[\sigma_2^{-4}-10\sigma_2^{-2}]\nonumber\\
&\hspace{18mm}
+[(3 \epsilon-2 \delta)\sigma_2^{-4}
+(-9 \epsilon+8 \delta)\sigma_2^{-2}]
+\Order_{\rm SR}(2;\sigma_2)\Big\}\;\text{(intermediate UV)}.
\label{eq:Csw00}
\end{align}
\endgroup
\begingroup
\begin{align}
\widehat\tau_{ii}^{\mathrm{SW}}={}&\frac{2 \pi C_k G}{3 a^{2} \epsilon}\Big\{k^{4}[1+(-\epsilon-2 \delta)+(2 \delta^{2})]\nonumber\\
&\hspace{18mm}
+k^{2}\cH^{2}[10+(-\epsilon+4 \delta)+(-11 \epsilon^{2}
+22 \delta \epsilon-14 \delta^{2})+(3 \epsilon \delta^{2})]\Big\}
\nonumber\\
={}&\frac{2 \pi C_k G \cH^{4}}{3 a^{2} \epsilon}\Big\{[1+(-\epsilon-2 \delta)+(2 \delta^{2})]\sigma_2^{-4}
+[10+(-\epsilon+4 \delta)+(-11 \epsilon^{2}
+22 \delta \epsilon-14 \delta^{2})\nonumber\\
&\hspace{18mm}
+(3 \epsilon \delta^{2})]\sigma_2^{-2}\Big\}\;\text{(strict UV)}
\nonumber\\
\simeq{}&\frac{2 \pi C_k G \cH^{4}}{3 a^{2} \epsilon}\Big\{[\sigma_2^{-4}
+10\sigma_2^{-2}]\nonumber\\
&\hspace{18mm}
+[(-\epsilon-2 \delta)\sigma_2^{-4}
+(-\epsilon+4 \delta)\sigma_2^{-2}]
+\Order_{\rm SR}(2;\sigma_2)\Big\}\;\text{(intermediate UV)}.
\label{eq:Cswii}
\end{align}
\endgroup
The strict UV limit is $(1,1/3)\factSW/\epsilon$. Thus $w_{\rm raw}\simeq1/3$, with the common $1/\epsilon$ enhancement. The
intermediate expressions retain the distinct finite-wavelength corrections.

\end{document}